\documentclass[a4paper,12pt,draft,aps,superscriptaddress,nofootinbib]{revtex4-2}

\usepackage{latexsym,amssymb,amsmath,mathrsfs,bbold,bm} 
\usepackage[final]{graphicx} 
\usepackage{mathtools}
\usepackage{physics}
\usepackage{slashed}
\usepackage{empheq}
\usepackage[utf8]{inputenc}
\usepackage{layout}

\usepackage{tikz}
\usetikzlibrary{decorations.pathreplacing}

\usepackage[a4paper, left=1.5cm, right=1.5cm, top=2.0cm, bottom=2cm]{geometry}
\newcommand \be{\begin{equation}}
\newcommand \ee{\end{equation}}
\newcommand \bea{\begin{eqnarray}}
\newcommand \eea{\end{eqnarray}}

\newcommand{\ri}{{\rm i}}
\newcommand\LFD[1]{\frac{\overrightarrow{\delta }}{\delta #1}}
\newcommand\RFD[1]{\frac{\overleftarrow{\delta }}{ \delta #1}}

\begin{document}
\allowdisplaybreaks

\title{Functional renormalization group study of the quark-meson model beyond the local potential approximation: Kurganov--Tadmor versus the grid method}
\author{Zsolt Sz{\'e}p}
\email{szepzs@achilles.elte.hu}
\affiliation{HUN-REN--ELTE Theoretical Physics Research Group, 1117 Budapest, Hungary}
\affiliation{Theory Department, HUN-REN Wigner RCP, POB 49, Budapest, 1525, Hungary}
\author{György Wolf}
\email{wolf.gyorgy@wigner.hu}
\affiliation{Theory Department, HUN-REN Wigner RCP, POB 49, Budapest, 1525, Hungary}

\begin{abstract}
We solve the functional renormalization group flow of the two-flavor quark-meson model at finite temperature and quark chemical potential with Litim's regulator, using the Kurganov--Tadmor central scheme and the standard grid method, both in the local potential approximation (LPA) and in an LPA$'$ truncation with running wave function renormalization factors for the pion, sigma, and quark fields.
Using the Kurganov--Tadmor scheme as a high-resolution benchmark down to very low temperatures, where the Fermi surface produces the sharpest fronts in the field derivative of the potential, we find that, with sufficiently fine resolution, the grid method gives virtually identical results in both truncations.
We discuss the difficulties that the wave function renormalization factors bring into the parametrization of the model and into the flow at finite density, related to the choice of the expansion point.
Contrary to some expectations, the peculiar backbending of the first-order phase boundary at low temperature, known from LPA, persists in our implementation of the LPA$'$ truncation, both with a fixed and with a running expansion point.
\end{abstract}

\maketitle

\section{Introduction}

The phase structure of strongly interacting matter at low temperature and large chemical potential, which is inaccessible to lattice QCD because of the sign problem \cite{Aarts:2023vsf}, is commonly explored with effective models such as the quark-meson model, treated beyond mean field with the functional renormalization group (FRG) based on the Wetterich equation \cite{Wetterich:1992yh}.

The two-flavor quark-meson (QM) model has been extensively investigated in the literature in various approximations of the derivative expansion \cite{Jungnickel:1995fp, Berges:1997eu, Tripolt:2013jra, Pawlowski:2014zaa, Chen:2021iuo}.
Such investigations also include studies \cite{Schaefer:1999em, Schaefer:2004en} based on the proper-time flow equation (derived via RG improvements of the one-loop effective action using Schwinger's proper-time regularization), which was proven equivalent \cite{Litim:2001hk, Litim:2001up} to an optimized Wetterich flow equation in the local potential approximation (LPA).
Results obtained with the proper-time flow in LPA can therefore be compared directly with those obtained using Litim's regulator.

Since the Wetterich flow equation at leading order in the derivative expansion maps onto an advection--diffusion-type partial differential equation, recent studies have proposed solving it by adapting and implementing advanced numerical techniques from computational fluid dynamics~\cite{Toro2009}.
Concrete applications to FRG flow equations include the discontinuous Galerkin method \cite{Grossi:2019urj}, as well as high-resolution finite-volume schemes such as the Kurganov--Tadmor (KT) method, discussed in a field-theoretical context in Ref.~\cite{KoenigsteinSeries} and recently applied to the QM model in LPA in Ref.~\cite{Stoll:2026ezu}.
These methods are particularly well suited to situations where the effective potential varies rapidly over a narrow interval of the field variable.
Such behavior can arise at small RG scales due to the flattening of the potential (as the effective potential needs to become convex) and, in theories with fermions, near the field value at which the quark energy $E_{\psi,k}(\phi)$ equals the chemical potential, where the Fermi--Dirac distribution approaches a step function in the zero-temperature limit.
In these regions the solution develops sharp fronts, analogous to shock waves in fluid dynamics.
When the potential is discretized on a fixed grid, such fronts may eventually become much narrower than the grid spacing, and conventional discretization schemes may then struggle to resolve them.

A peculiarity of the quark-meson model in the FRG is that in LPA with Litim's regulator a liquid--gas type of transition appears at low temperature, separating two phases with different nonzero condensate values, corresponding to a broken symmetry phase and a dense phase.
The low-temperature part of the phase-transition line in the $T$--$\mu$ plane bends towards the origin, $dT_{\rm c}/d\mu_{\rm c}>0$ (backbending), in contrast with the standard mean-field result~\cite{Scavenius:2000qd}, where a chiral transition takes place and the boundary line satisfies $dT_{\rm c}/d\mu_{\rm c}<0$.
These features were observed in the chiral limit in Ref.~\cite{Schaefer:2004en} and at the physical pion mass in Ref.~\cite{Tripolt:2017zgc}, where their physical implications were studied, in particular a negative entropy density of the dense phase.

Subsequent studies addressed the question of whether the backbending is a genuine property of the model or an artifact of the approximations involved: the treatment of the field dependence of the effective potential, the order of the derivative expansion, the numerical method, or the regulator.
Taylor expansions of the effective potential about its scale-dependent minimum turned out to be of limited use: even at higher orders they do not resolve the low-temperature part of the first-order line, where the second minimum lies beyond their radius of convergence~\cite{Pawlowski:2014zaa}, while at low order they produce neither the backbending nor a first-order transition (see Fig.~10 in Ref.~\cite{Zhang:2017icm}, chiral limit).
Using the grid method, the absence of backbending was reported in Sec.~4.2 of the Ph.D.\ thesis~\cite{Weyrich:2018jid} in an $\mathrm{LPA}'$ truncation, which goes beyond LPA by including scale-dependent wave function renormalizations: two independent ones for the two mesonic degrees of freedom, and one for the quarks, which was found to be of minor importance.
With a numerical method based on a Chebyshev expansion of the effective potential, the backbending was reproduced in LPA but not seen in the $\mathrm{LPA}'$ approximation used in Ref.~\cite{Chen:2021iuo}, which incorporates a scale- and field-dependent mesonic wave function renormalization.
In this approximation the first-order line was computed only in the chiral limit and down to $T\approx30$~MeV, i.e., above the temperature at which the line bends back in LPA (see the insets of Fig.~2 there), since the numerics becomes difficult at low $T$ and large $\mu$.
It was conjectured in Ref.~\cite{Chen:2021iuo} that the backbending could arise from the discontinuities in the regime of the first-order transition, which are not fully captured by standard methods used to compute the global effective potential, such as the grid method and the Chebyshev expansion.
However, the discontinuous Galerkin method, which allows for discontinuities in the flow, still yields the backbending in LPA in the chiral limit (see Fig.~8 in Ref.~\cite{Grossi:2021ksl}).
The role of the regulator was investigated in Ref.~\cite{Otto:2022jzl}, which showed that for Callan--Symanzik-type regulators the backbending disappears and, at the physical pion mass, the transition becomes a smooth crossover down to $T=0$.
This study traced the backbending obtained with Litim's regulator back to the sharp decoupling of the fermions at the Fermi surface.
Most recently, Ref.~\cite{Stoll:2026ezu} solved the LPA flow at the physical pion mass with the KT scheme for three spatial momentum regulators of different analytic properties, including Litim's regulator, for UV cutoffs in the range $0.6$--$1.2$~GeV, and computed the phase diagram for $T>10$~MeV.
The backbending was found in all cases, although less pronounced for the two smooth regulators, and was therefore concluded to be a feature of the LPA truncation rather than a regulator or UV-cutoff artifact, which might disappear at higher orders of the derivative expansion.
Taken together, these results indicate that the backbending is not a failure of the numerical methods used to solve the flow equation, but a consequence of the truncation of the Wetterich equation in combination with the choice of the regulator.
Its fate beyond LPA remains to be investigated.

In view of the above summary of existing results, our objectives in this paper are threefold.
First, we implement the KT scheme for the two-flavor QM model in LPA and test how it performs at low temperature near the Fermi surface.
Second, we use the KT results as a high-resolution benchmark to assess the accuracy of the grid method in this region, at temperatures well below those for which the phase diagram was computed in Ref.~\cite{Stoll:2026ezu}, where the fronts are the sharpest.
To our knowledge, such a comparison has not been performed for this model so far; it is also a prerequisite for using the grid method beyond LPA.
Finally, we use both numerical methods in the $\mathrm{LPA}'$ truncation of Ref.~\cite{Weyrich:2018jid} containing three running wave function renormalization factors to investigate whether the backbending persists beyond LPA or is absent, as reported there using Litim's regulator.

The paper is organized as follows.
After introducing the model, its LPA$'$ truncation, and the definitions of the wave function renormalization factors in Sec.~\ref{sec:FRG}, we describe in Sec.~\ref{sec:KT} how the flow equation is cast in advection--diffusion form and solved with the KT scheme.
The parameters of the model are fixed in Sec.~\ref{sec:param}.
The LPA results, including the comparison between the KT and grid methods, are presented in Sec.~\ref{sec:LPA}, while Sec.~\ref{sec:LPAp} contains the flow equations of the $Z$ factors and the LPA$'$ phase diagrams obtained with a fixed and a running expansion point.
We summarize and conclude in Sec.~\ref{sec:concl}.
The flow equations of the two-point functions are given in Appendix~\ref{App:2point}, and the method used to evaluate their $\mathcal{O}(\vec p^{\,2})$ contributions is presented in Appendix~\ref{app:example}.

\section{The FRG approach in the quark-meson model}\label{sec:FRG}

In a theory containing interacting bosonic and fermionic fields, the scale-dependent effective average action is determined by the flow equation \cite{Wetterich:1992yh}
\begin{equation}
  \partial_k \Gamma_k[\Phi] = \frac{1}{2}\textrm{STr} \Big[ \big(\Gamma_k^{(1,1)}[\Phi] + {\bf R}_k\big)^{-1} \partial_k {\bf R}_k\Big],
  \label{Eq:_Gamma_compact}
\end{equation}
where ${\bf R}_k$ is the regulator matrix and $\Gamma_k^{(1,1)}[\Phi]$ is the second functional derivative of the effective average action with respect to the multicomponent boson and fermion background fields collected in the vector $\Phi^{\rm T}=(\vec\phi^{\, \rm T},\psi^{\rm T},\bar \psi)$. 
$\textrm{STr}$ contains, in addition to a functional integral, a summation over color, flavor and Dirac indices and, due to their Grassmann nature, a minus sign for fermions.
As discussed in Ref.~\cite{Berges:2000ew}, evaluated on a purely bosonic background, $\Gamma_k^{(1,1)}[\Phi]$ becomes block-diagonal: $\Gamma_k^{(1,1)}=\textrm{diag}(\Gamma_{k,{\rm B}}^{(2)},\Gamma_{k,{\rm F}}^{(1,1)})$, where the bosonic and fermionic blocks are denoted by indices ${\rm B}$ and ${\rm F}$.
With a regulator matrix of the same form, ${\bf R}_k=\textrm{diag}({\bf R}_k^{\rm B},{\bf R}_k^{\rm F})$, the right-hand side of the Wetterich flow equation becomes the difference of bosonic and fermionic contributions:
\begin{equation}
  \partial_k \Gamma_k[\Phi] = \frac{1}{2}\textrm{Tr} \Bigg[ \bigg(\frac{\delta^2\Gamma_k[\Phi]}{\delta\vec\phi^{\rm T}\delta\vec\phi} + {\bf R}_k^{\rm B}\bigg)^{-1} \partial_k {\bf R}_k^{\rm B}\Bigg] - \textrm{Tr} \Bigg[\bigg(\LFD{\Psi^{\rm T}}\Gamma_k[\Phi] \RFD{\Psi} + {\bf R}_k^{\rm F}\bigg)^{-1} \partial_k {\bf R}_k^{\rm F}\Bigg],
  \label{Eq:_Gamma_B-F}
\end{equation}
where $\Psi^{\rm T}(-P)=\big(\psi^{\rm T}(-P),\bar\psi(P)\big)$.
In order to solve this equation in a particular model, a physically motivated ansatz for the effective average action and an initial functional at some UV scale $\Lambda$ are needed.
The field-theoretical effective action is obtained by integrating the flow equation down to $k=0$, in practice to some appropriate IR scale $k_{\rm IR}$.

\subsection{QM model in the LPA$'$ truncation}

The quark-meson model describes the interaction of $N_f = 2$ constituent quark flavors with a chiral meson multiplet $\vec{\phi}^\textrm{T}=(\sigma,\bm{\pi})$ that forms the fundamental representation of the $\mathrm{O}(N=4)$ symmetry group, with $\mathrm{SO}(4)\simeq \mathrm{SU}(2)_L \times \mathrm{SU}(2)_R$ locally.
For $\Gamma_k$ we adopt the ansatz used in Ref.~\cite{Weyrich:2018jid}, since we would like to check the LPA$'$ results reported there:
\bea
\Gamma_k[\Phi] &=& \int_x \Bigl[
    Z_{\psi,k}\bar \psi\,\tilde{\slashed{\partial}}\,\psi + g\,\bar \psi\bigl(\sigma\mathbb{1} - \ri \gamma_5\bm{\tau}\cdot\bm{\pi}\bigr)\psi
    + \frac{Z_{\phi,k}}{2}\Big(\big(\partial_\tau\phi_a\big)^2 + \big(\bm{\nabla}\phi_a\big)^2 \Big) \nonumber \\
    &&\qquad + \frac{Y_k}{4} \Big(\big(\partial_\tau\rho\big)^2 + \big(\bm{\nabla}\rho\big)^2 \Big) + U_k(\rho) - h_a\,\phi_a
  \Bigr].
  \label{Eq:ansatz}
\eea
in which the term $Y_k$ multiplying the derivatives of the $\mathrm{O}(N)$-invariant $\rho=\phi_a\phi_a/2$ allows for different wave function renormalizations of the sigma and pion fields in the broken phase, while keeping the ansatz $\mathrm{O}(N)$ invariant.
The last term in Eq.~\eqref{Eq:ansatz} is the explicit chiral-symmetry breaking source, while the effective potential $U_k(\rho)$ depends on chiral invariants \cite{Jungnickel:1995fp,Berges:1997eu}.
We work in the imaginary-time formalism of finite-temperature field theory, and in Eq.~(ansatz) we used the shorthands $\int_x:= \int_0^{1/T} d\tau \int d^d x$ and $\tilde{\slashed{\partial}}:=\gamma^0 (\partial_\tau -\mu) - \ri \bm{\gamma}\cdot\bm{\nabla}$, where $T$ is the temperature, $\mu$ is the quark chemical potential and the gamma matrices satisfy $\{\gamma^\mu,\gamma^\nu\}=2g^{\mu\nu}$ with $g^{\mu\nu}$ being the Minkowski metric with signature $(1,-1,-1,-1)$.

The presence of $Z_{\phi,k}$ and $Y_k$ leads to separate wave function renormalization factors for the pion and sigma fields:
\be
Z_{\pi,k}=Z_{\phi,k},\qquad Z_{\sigma,k}=Z_{\phi,k} + \rho_\star Y_k.
\ee
Here $\rho_\star=\phi_\star^2/2$ plays the role of the expansion point of the derivative expansion: in general, the coefficients of the kinetic terms depend on the field and on the momentum, $Z_{\phi,k}(\rho,-\partial^2)$ and $Y_k(\rho,-\partial^2)$, and the ansatz \eqref{Eq:ansatz} retains only their values at $\rho=\rho_\star$ and vanishing momentum \cite{Wetterich:1991be}.
The $Z$ factors thus obtained are field independent and are used for all values of the field in the flow of the effective potential.
Usually, $\rho_\star$ is taken to be the running minimum of the potential, but it can also be kept fixed at some value.
When $\rho_\star$ is the running minimum, we neglect in the flow of $Z_{\sigma,k}$ the contribution $Y_k\,\partial_k\rho_\star$ arising from the $k$ dependence of the minimum.

On the right-hand side of Eq.~\eqref{Eq:_Gamma_B-F}, the functional derivatives $\Gamma_{k,{\rm B}}^{(2)}$ and $\Gamma_{k,{\rm F}}^{(1,1)}$ are evaluated on a homogeneous ($x$-independent) mesonic background and a vanishing fermionic one, with a bosonic regulator that is chosen diagonal in flavor space.
We use the 3D version of the LPA-optimized regulator functions \cite{Litim:2001up,Litim:2000ci}
\begin{equation}
\begin{aligned}
  {\bf R}^{\rm B}_{k}(P,-Q) &= (2\pi)^{D}\delta^{(D)}(P-Q)\,\vec{p}^{\,2} r_k^{\rm B}(\vec{p}^{\,2}/k^2)
  \begin{pmatrix}
    Z_{\sigma,k} & 0 \\ 0 & Z_{\pi,k}\mathbb{1}_{N-1}
  \end{pmatrix},\\
  {\bf R}_k^{\rm F}(P,-Q) &= (2\pi)^{D}\delta^{(D)}(P-Q)\, Z_{\psi,k} r^\textrm{F}_k(\vec{p}^{\,2}/k^2)
   \begin{pmatrix}
    0 & \vec{\slashed{p}}^{\rm T}  \\
    \vec{\slashed{p}} & 0
  \end{pmatrix},
  \end{aligned}
\end{equation}
where $D=d+1$ is the dimension of Euclidean space-time (at finite temperature $(2\pi)^{D}\delta^{(D)}(P-Q)$ is understood as $T^{-1}\delta_{n_P n_Q}(2\pi)^{d}\delta^{(d)}(\vec p-\vec q)$, with $n_P$ and $n_Q$ labeling the Matsubara frequencies), and the dimensionless shape functions are given by $r^\textrm{B}_k(x)=\big(x^{-1}-1\big)\Theta(1-x)$ and $r^\textrm{F}_k(x)=\big(x^{-1/2}-1\big)\Theta(1-x)$.
We mention that the bosonic part of the regulator matrix, adopted from Ref.~\cite{Weyrich:2018jid} has to be taken with a grain of salt, as for $Y_k\ne0$ it breaks the $O(4)$ invariance at the regulator level.

Being linear in the field, the explicit symmetry-breaking term does not contribute to $\Gamma_{k,{\rm B}}^{(2)}$.
As a result, it does not contribute to the pion and sigma curvature masses, which are the eigenvalues of the curvature mass matrix $M^2_{ij,k}=\frac{\partial^2 U_k}{\partial\phi_i\partial\phi_j}$.
Taking derivatives with respect to the $\mathrm{O}(N)$-invariant $\rho=\phi_i\phi_i/2$, one can write
\begin{equation}
  M^2_{ij,k}(\rho) = \delta_{ij} U'_k(\rho) + \phi_i \phi_j U''_k(\rho) = M_{\pi,k}^2 \mathbb{P}^{\mathrm{T}}_{ij} + M_{\sigma,k}^2 \mathbb{P}^{\mathrm{L}}_{ij},
  \label{Eq:hat_M2}
\end{equation}
where $\mathbb{P}^{\mathrm{T}}=\mathbb{1} - \mathbb{P}^{\mathrm{L}}$ and $\mathbb{P}^{\mathrm{L}}_{ij}=\frac{\phi_i \phi_j}{\phi^2}$ are the transverse and longitudinal projectors, and
\begin{subequations}
\begin{equation}
  M_{\pi,k}^2(\rho) = U'_k(\rho), \qquad M_{\sigma,k}^2(\rho) = M_{\pi,k}^2(\rho) + 2 \rho U''_k(\rho)
\label{Eq:hat_M2_pi_sig}
\end{equation}
are the squares of the \emph{running} curvature masses.

The form of the curvature masses given in Eq.~\eqref{Eq:hat_M2_pi_sig} emphasizes that the flow equation depends solely on $\mathrm{O}(N)$-invariant quantities.
Since it is often more convenient to work on a $\sigma$-grid rather than a $\rho$-grid, one defines $\mathcal{U}_k(\sigma) \equiv U_k(\rho = \sigma^2/2)$.
Choosing the external field along the $\sigma$ direction, $\vec{h} = (h, \vec{0})$, and evaluating the curvature masses on a background field $\vec{\phi} = (\sigma, \vec{0})$, yields:
\begin{equation}
M_{\pi,k}^2(\sigma) = \frac{1}{\sigma} \frac{\partial \mathcal{U}_k(\sigma)}{\partial \sigma}, \qquad
M_{\sigma,k}^2(\sigma) = \frac{\partial^2 \mathcal{U}_k(\sigma)}{\partial \sigma^2}. 
\label{Eq:Ms2_spec_bg}
\end{equation}
\end{subequations}

With these definitions in place, one can formulate the flow equation for $\mathcal{U}_k(\sigma)$ at finite temperature $T$ and quark chemical potential $\mu$.
A detailed derivation using standard finite-temperature functional techniques (see, e.g., Sec.~2.4 of Ref.~\cite{Almasi:2017lso} for the LPA case) yields
\begin{equation}
\partial_k \mathcal{U}_k(\sigma;\mu,T) = \frac{k^{d+1}\Omega_d}{2 d} \Bigg[ \sum_{i=\sigma,\pi} \frac{c_i}{E_{i,k}} \coth\left(\frac{E_{i,k}}{2T}\right) - \sum_{t=\pm} \frac{c_\psi}{E_{\psi,k}} \tanh\left(\frac{E_{\psi,k} - t\mu}{2T}\right) \Bigg],
\label{Eq:T_flow}
\end{equation}
where $\Omega_d = \big[2^{d-1}\pi^{d/2}\Gamma\left(\frac{d}{2}\right)\big]^{-1}$, $c_\sigma = \big(1-\frac{\eta_{\sigma}}{d+2}\big)$, $c_\pi = 3\big(1-\frac{\eta_{\pi}}{d+2}\big)$, $c_\psi=2 N_c N_f \big(1-\frac{\eta_{\psi}}{d+1}\big)$, with $N_c = 3$, and $N_f = 2$.
For the fermionic modes, $t = +1$ ($-1$) labels the particle (antiparticle) contribution and the energies are defined as
\bea
E^2_{\pi,k} = k^2 + \frac{M^2_{\pi,k}}{Z_{\pi,k}},\quad
E^2_{\sigma,k} = k^2 + \frac{M^2_{\sigma,k}}{Z_{\sigma,k}}, \quad
E^2_{\psi,k} = k^2 + \frac{M^2_q}{Z^2_{\psi,k}},
\eea
where $M_q = g \sigma$ is $k$-independent, while the anomalous dimensions are introduced through the relation ($i=\pi,\sigma,\psi$)
\be
\eta_i = -\frac{k\partial_k Z_{i,k}}{Z_{i,k}}\,.
\label{Eq:eta_def}
\ee

Physically, the negative quark-loop contribution favors chiral symmetry breaking by driving the formation of a non-trivial minimum at non-zero field values. 
In contrast, the mesonic loop contributions oppose symmetry breaking and, in the infrared ($k \to 0$), dominate the flow within the region of negative curvature ($M_{\sigma,k}^2 < 0$). 
There, soft sigma fluctuations drive the flattening of the inner potential barrier for small field values, thereby restoring the convexity of the effective action \cite{Tetradis:1992qt, Litim:2000ci}.

At zero temperature and finite chemical potential, the flow equation reduces to
\begin{equation}
\partial_k \mathcal{U}_k(\sigma;\mu,0) = \frac{k^{d+1}\Omega_d}{2d} \left[ \frac{c_\sigma}{E_{\sigma,k}} + \frac{c_\pi}{E_{\pi,k}} \right] - \frac{k^{d+1}\Omega_d}{d} \frac{c_\psi}{E_{\psi,k}} \, \Theta\bigl(E_{\psi,k} - \mu\bigr),
\label{Eq:T0_flow}
\end{equation}
where the Heaviside step function gives the high-energy part of the fermionic vacuum fluctuations.
This follows from the identity $\Theta(E_{\psi,k} - \mu) = 1 - \Theta(\mu - E_{\psi,k})$, in which the vacuum contribution ($1$) is partially canceled by the matter contribution $\Theta(\mu - E_{\psi,k})$ whenever the Fermi sphere is occupied ($E_{\psi,k} < \mu$).
Therefore, fermions only contribute to the flow of the potential outside the Fermi sphere.

\subsection{Wave function renormalization}

The flow equations of the wave function renormalization factors follow from those of the two-point functions.
Taking two functional derivatives of Eq.~\eqref{Eq:_Gamma_compact} with respect to the fields $\Phi_a$ and $\Phi_b$ one obtains
\begin{equation}
    \partial_k \Gamma^{(2)}_{k,ab} = \frac{1}{2}\tilde\partial_k\,\textrm{STr}\Big[{\bf G}_k\,\Gamma^{(4)}_{k,ab}\Big]
  -\frac{1}{2}\tilde\partial_k\,\textrm{STr}\Big[{\bf G}_k\,\Gamma^{(3)}_{k,a}\,{\bf G}_k\,\Gamma^{(3)}_{k,b}\Big],
  \label{Eq:Gamma2_flow}
\end{equation}
where ${\bf G}_k=\big(\Gamma_k^{(1,1)}+{\bf R}_k\big)^{-1}$ is the regularized propagator, $\tilde\partial_k$ acts only on the $k$ dependence of the regulator, i.e., $\tilde\partial_k{\bf G}_k=-{\bf G}_k\,\partial_k{\bf R}_k\,{\bf G}_k$, while $\Gamma^{(3)}_{k,a}$ and $\Gamma^{(4)}_{k,ab}$ denote the first and second functional derivatives of $\Gamma_k^{(1,1)}$ with respect to $\Phi_a$ and $\Phi_a,\Phi_b$ (for fermionic $\Phi_a$, $\Phi_b$ the Grassmann signs have to be taken into account).
The first term corresponds to the tadpole diagram with a four-point vertex, the second one to the one-loop diagrams with two three-point vertices.
In the ansatz \eqref{Eq:ansatz}, the four-point vertices depend on the external momentum only through the terms proportional to $Y_k$; hence, the tadpole contributes to the flow of the bosonic wave function renormalizations only through $Y_k$.
The explicit form of Eq.~\eqref{Eq:Gamma2_flow} in momentum space, together with the vertices entering the flow of the sigma and pion two-point functions, is given in Eq.~\eqref{Eq:2point_flow_mom} of Appendix~\ref{App:2point}.

Since the regulator acts only on the spatial momenta, the wave function renormalizations are defined from the dependence of the two-point functions on the spatial external momentum $\vec p$, with $P=(P^0,\vec p)$.
For the mesons one uses the usual definition
\begin{equation}
  Z_{\pi,k}=\lim_{P\to P^{\rm B}_\textrm{low}}\frac{\partial}{\partial \vec p^{\,2}}\,\Gamma^{(2)}_{k,\pi\pi}(P)\bigg|_{\sigma=\phi_\star},\qquad
  Z_{\sigma,k}=\lim_{P\to P^{\rm B}_\textrm{low}}\frac{\partial}{\partial \vec p^{\,2}}\,\Gamma^{(2)}_{k,\sigma\sigma}(P)\bigg|_{\sigma=\phi_\star},
  \label{Eq:Zb_def}
\end{equation}
with $P^{\rm B}_\textrm{low}=(0,\vec 0)$, which, applied to the ansatz \eqref{Eq:ansatz}, reproduces $Z_{\pi,k}=Z_{\phi,k}$ and $Z_{\sigma,k}=Z_{\phi,k}+\rho_\star Y_k$.
For the quarks, the wave function renormalization is projected out from the part of the two-point function proportional to $\vec{\slashed{p}}=\vec\gamma\cdot\vec p$:
\begin{equation}
  Z_{\psi,k}=-\frac{1}{4N_cN_f}\,\mathfrak{R}\lim_{P\to P^{\rm F}_\textrm{low}}\frac{\partial}{\partial \vec p^{\,2}}\,
  \textrm{tr}\Big[\vec{\slashed{p}}\;\Gamma^{(2)}_{k,\bar\psi\psi}(P)\Big]\bigg|_{\sigma=\phi_\star},
  \label{Eq:Zf_def}
\end{equation}
where the trace is taken over Dirac, color and flavor indices, and the minus sign compensates $\textrm{tr}\big[\mathbb{1}\vec{\slashed{p}}\,\vec{\slashed{p}}\big]=-4N_cN_f\,\vec p^{\,2}$ for Minkowski gamma matrices.
Since zero is not a fermionic Matsubara frequency, the limit is taken towards the lowest one, $P^{\rm F}_\textrm{low}=(P^0_\textrm{low},\vec 0)$ with $P^0_\textrm{low}=\ri\pi T-\mu$, which includes the chemical potential, and, as is customary in the literature \cite{Friederich:2010zn, Tanizaki:2013doa, Pawlowski:2014zaa}, the real part is taken.
The flow equations of the wave function renormalizations are then obtained by applying the projections \eqref{Eq:Zb_def} and \eqref{Eq:Zf_def} to the right-hand side of Eq.~\eqref{Eq:Gamma2_flow}: the flows of $Z_{\pi,k}$ and $Z_{\sigma,k}$ receive contributions from the mesonic loops (the $\pi\sigma$ loop for the pion, the $\pi\pi$ and $\sigma\sigma$ loops for the sigma) and from the quark loop, while the flow of $Z_{\psi,k}$ is given by the loops containing one quark and one pion or sigma propagator.
Since the regulators are proportional to the corresponding $Z$ factors, $\partial_k{\bf R}_k$ contains the anomalous dimensions, which therefore appear on both sides of the flow equations.
This leads to a linear system of equations for $\eta_\pi$, $\eta_\sigma$, and $\eta_\psi$, which is easily obtained from Eqs.~\eqref{Eq:Zpi_flow}, \eqref{Eq:Zsigma_flow}, and \eqref{Eq:Zpsi_flow} and has to be solved at each RG scale $k$.

\subsection{Minimum of the effective potential and physical masses}

To obtain the physical masses, one evaluates the curvature masses at the minimum $v$ of the potential obtained at $k_{\mathrm{IR}}$ and divides them by the corresponding wave function renormalization factors: $m_\pi^2=M^2_{\pi,k_{\mathrm{IR}}}(v)/Z_{\pi,k_{\mathrm{IR}}}$ and $m_\sigma^2=M^2_{\sigma,k_{\mathrm{IR}}}(v)/Z_{\sigma,k_{\mathrm{IR}}}$, while the constituent quark mass is $g v/Z_{\psi,k_{\mathrm{IR}}}$.
In the case of explicit symmetry breaking, the extremum is determined from the equation $\mathcal{U}'_k(v) - h = 0$. 
Therefore, $h$ enters directly into the expression for the pion curvature mass, $M_{\pi,k}^2(v) = h/v$, as dictated by the axial Ward identity, and indirectly into the sigma curvature mass through the value of $v$. 
In terms of the renormalized minimum $\bar\phi_R=Z^{1/2}_{\pi,k_{\mathrm{IR}}}\,v$ one has $m_\pi^2=h/\big(Z^{1/2}_{\pi,k_{\mathrm{IR}}}\,\bar\phi_R\big)$.
Choosing $h = Z^{1/2}_{\pi,k_{\mathrm{IR}}}\,f_\pi m_\pi^2$ when parameterizing the model at $T = 0$, with $m_\pi$ being the physical mass, therefore guarantees that the pion mass takes its physical value, provided that at $k_{\mathrm{IR}}$ the global minimum satisfies $\bar\phi_R^{T=0} = f_\pi$.
Since $Z_{\pi,k_{\mathrm{IR}}}$ is known only after the flow has been integrated, $h$ has to be determined together with the other parameters of the model.
In LPA, where all the $Z$ factors are equal to one, these conditions reduce to $h = f_\pi m_\pi^2$ and $v^{T=0} = f_\pi$.

\section{Numerical solution of the flow equation in the KT scheme}\label{sec:KT}

\subsection{The flow equation in the advection--diffusion form}

Taking the spatial derivative of both sides in Eq.~\eqref{Eq:T_flow} and introducing the notation $s \equiv \sigma$, the scale evolution equation for $u(s,k) \equiv \partial_s \mathcal{U}_k$ takes the non-linear advection--diffusion form \cite{Toro2009} 
\begin{equation}
  \partial_k u(s, k) + \frac{d}{ds} F(s,u,k) = \frac{d}{ds} Q(u_s,k) + S(s,k)\,.
  \label{Eq:T-flow_ADF}
\end{equation}
Here, $d/ds$ denotes the total spatial derivative at fixed $k$, accounting for both explicit $s$-dependence in $F$ and spatial variations of the field $u(s,k)$. Partial derivatives denote differentiation with respect to individual arguments, e.g., $u_s \equiv \partial u/\partial s$.

The second term on the left-hand side, $dF/ds$, represents the advective derivative of the pion flux, capturing the non-linear, conservative transport of pion degrees of freedom across scale $k$ and field coordinate $s$.
Here, $F$ depends on $s$ through the ratio $u/s$, which corresponds to the running pion curvature mass $M^2_{\pi,k}$ in Eq.~\eqref{Eq:Ms2_spec_bg}.
On the right-hand side, $dQ/ds$ represents the spatial derivative of the sigma flux, acting as a non-linear diffusive dissipation driven by field gradients $u_s$.
The second term, $S(s,k)$, is an explicit source accounting for non-conservative contributions from quark fluctuations.
In constructing $S(s,k)$, the total and partial $s$-derivatives coincide because the quark energy $E_{\psi,k}$ depends on $s$ only explicitly.
Note that Eq.~\eqref{Eq:T-flow_ADF} is simpler than the general expression in Ref.~\cite{KoenigsteinSeries} [Eq.~(109)], as $S$ and $Q$ do not depend on $u$, and $Q$ has no explicit $s$-dependence.
In LPA$'$ the $Z$ factors and anomalous dimensions enter $F$, $Q$, and $S$ as field-independent coefficients, kept fixed during each RG step (see Sec.~\ref{ss:Zflow}).
We also tested an implementation where the quark contribution was included in the flux rather than as a source term, and we obtained identical results.

By examining the flow equation~\eqref{Eq:T0_flow} for the potential, one observes that the corresponding zero-temperature flow equation for $u(s,k)$ contains a Dirac delta singularity at the Fermi surface (edge), originating from the field derivative of the step function, $\partial_s \Theta(E_{\psi,k} - \mu) = \delta(E_{\psi,k} - \mu) \, \partial_s E_{\psi,k}$. 
This sharp behavior makes the equation difficult to solve. 
Our goal is therefore to construct a stable numerical solution for the non-linear advection--diffusion system~\eqref{Eq:T-flow_ADF} capable of handling sharp gradients over the spatial domain $[s_{\min}, s_{\max}]$, starting from the ultraviolet initial condition at scale $k = \Lambda$ and respecting the physical boundary conditions to be prescribed at $s_{\min}$ and $s_{\max}$.

\subsection{The finite-volume method}

The \emph{finite-volume} (FV) method transforms partial differential equations written in conservation form into a system of discrete algebraic equations for the cell-averaged state variables.
By integrating the differential equation over each individual grid cell, that is the small subregions that partition the computational mesh, spatial integrals are converted via the divergence theorem into numerical fluxes evaluated at cell boundaries.
Because the flux leaving one cell is identically equal to that entering the adjacent cell, local conservation is guaranteed by construction.

In order to introduce some basic concepts of the finite-volume method, consider a one-dimensional non-linear conservation law whose flux, as in Eq.~\eqref{Eq:T-flow_ADF}, depends on the spatial coordinate not only implicitly, through the conserved variable, but also explicitly:
\begin{equation}
\frac{\partial u(s,k)}{\partial k} + \frac{d}{ds} F\big(s, u(s,k), k\big) = 0\,,
\label{Eq:cons_law}
\end{equation}
where $k$ is the evolution parameter, $s$ is the spatial coordinate, and $F(s,u,k)$ is the flux function.
The spatial derivative acts on both of its first two arguments,
\begin{equation}
  \frac{d}{ds} F\big(s,u(s,k),k\big) = \frac{\partial F}{\partial s} + \frac{\partial F}{\partial u}\, u_s\,,
  \label{Eq:dF_chain}
\end{equation}
and only the second, quasi-linear piece carries the characteristic speed, the first one acting as a geometric source.
To maintain strict flux conservation in the finite-volume scheme, both terms remain grouped inside the total derivative on the left-hand side rather than moving the geometric source term to the right-hand side.

Instead of solving the equation at discrete points, the spatial domain is divided into non-overlapping control volumes (cells) $I_i = [s_{i-1/2}, s_{i+1/2}]$ of uniform width $\Delta s = s_{i+1/2} - s_{i-1/2}$.
A standard discretization places the cell centers at $s_i = i \Delta s$ and the cell interfaces at
\begin{equation}
  s_{i+1/2} = \left(i + \frac{1}{2}\right)\Delta s = s_i + \frac{\Delta s}{2} = s_{i+1} - \frac{\Delta s}{2}\,,
  \label{Eq:interface_abscissa}
\end{equation}
where $i$ is an integer index.
This discretized interface abscissa is the value at which the explicit $s$-argument of the flux has to be evaluated below; it is displaced by half a cell from the cell centers $s_i$, $s_{i+1}$ at which the cell averages live.

Integrating the conservation law over the cell $I_i$ gives:
\begin{equation}
\int_{s_{i-1/2}}^{s_{i+1/2}} \frac{\partial u}{\partial k} \, ds + \int_{s_{i-1/2}}^{s_{i+1/2}} \frac{d}{ds} F\big(s,u(s,k),k\big) \, ds = 0\,.
\end{equation}
Applying the fundamental theorem of calculus transforms the spatial integral into exact boundary fluxes evaluated at the cell faces:
\begin{equation}
\frac{d}{dk} \bigg( \frac{1}{\Delta s} \int_{s_{i-1/2}}^{s_{i+1/2}} u(s,k) \, ds \bigg) + \frac{F\big(s_{i+1/2},u(s_{i+1/2}, k),k\big) - F\big(s_{i-1/2},u(s_{i-1/2}, k),k\big)}{\Delta s} = 0\,.
\end{equation}
Defining the \emph{cell average} $u_i(k)$ as:
\begin{equation}
u_i(k) \equiv \frac{1}{\Delta s} \int_{s_{i-1/2}}^{s_{i+1/2}} u(s,k) \, ds\,,
\end{equation}
the exact, semi-discrete finite-volume balance equation becomes:
\begin{equation}
\frac{d u_i(k)}{dk} = -\frac{F\big(s_{i+1/2},u(s_{i+1/2}, k),k\big) - F\big(s_{i-1/2},u(s_{i-1/2}, k),k\big)}{\Delta s}\,,
\label{Eq:FV_balance}
\end{equation}
where $F\big(s_{i\pm 1/2},u(s_{i\pm 1/2}, k),k\big)$ are the exact physical fluxes passing through the cell boundaries.
No approximation has been made so far: the fundamental theorem of calculus applies to the \emph{total} $s$-derivative \eqref{Eq:dF_chain}, so that the explicit $s$-dependence is accounted for exactly, simply by evaluating $F$ at the interface abscissas $s_{i\pm 1/2}$ of Eq.~\eqref{Eq:interface_abscissa}.
Moreover, since two neighboring cells share the same interface, they also share the same value of $s_{i+1/2}$ and hence one and the same flux, which enters their balance equations with opposite signs; the scheme therefore remains strictly conservative in spite of the explicit $s$-dependence of $F$.
In practice, because the exact point values $u(s_{i\pm 1/2}, k)$ are unknown, these are replaced by \emph{numerical fluxes} computed from reconstructed states.

Since the finite-volume scheme tracks and updates only cell-averaged values $u_i(k)$, point values $u(s, k)$ inside the cells are not explicitly known and must be reconstructed from surrounding cell averages $\{u_{i-1}, u_i, u_{i+1}, \dots\}$.
Regardless of whether the physical solution $u(s)$ is smooth or contains sharp features, independent cell-wise reconstruction creates an artificial \emph{jump discontinuity} at every cell interface $s_{i+1/2}$, yielding distinct left and right reconstructed states, $u^-_{i+1/2}$ and $u^+_{i+1/2}$.

The one-sided limits approaching the interface $s_{i+1/2}$ from the left and right are defined as 
\begin{equation}
  u_{i+1/2}^- \equiv \lim_{\epsilon \to 0^+} u(s_{i+1/2} - \epsilon)\,, \qquad
  u_{i+1/2}^+ \equiv \lim_{\epsilon \to 0^+} u(s_{i+1/2} + \epsilon)\,.
\end{equation}
To evaluate the interface values $u_{i+1/2}^\pm$ with higher accuracy than a flat cell average, we construct a local piecewise linear polynomial $L_i(s)$ inside each cell $I_i$:
\begin{equation}
L_i(s) = u_i + (s - s_i)(u_s)_i, \qquad s \in \left[s_{i-1/2}, \, s_{i+1/2}\right]\,,
\end{equation}
where $(u_s)_i$ is an approximate spatial slope inside cell $i$.
Evaluating $L_i(s)$ at $s = s_i + \frac{\Delta s}{2} = s_{i+1/2}$ and $L_{i+1}(s)$ at $s = s_{i+1} - \frac{\Delta s}{2} = s_{i+1/2}$ yields the left- and right-extrapolated interface states:
\begin{equation}
\begin{aligned}
  u_{i+1/2}^- &= L_i\big(s_{i+1/2}\big) = u_i + \frac{\Delta s}{2} (u_s)_i\,, \\
  u_{i+1/2}^+ &= L_{i+1}\big(s_{i+1/2}\big) = u_{i+1} - \frac{\Delta s}{2} (u_s)_{i+1}\,.
\end{aligned}
\end{equation}
Only the conserved variable is reconstructed in this way.
When evaluating fluxes at an interface, the pair $\big(s_{i+1/2}, u^\pm_{i+1/2}\big)$ must be used with $s_{i+1/2}$ given by the interface abscissa~\eqref{Eq:interface_abscissa}: replacing $s_{i+1/2}$ by the neighboring cell centers $s_i$ or $s_{i+1}$ would introduce an $\mathcal{O}(\Delta s)$ error in the first argument, destroying the second-order accuracy gained by reconstructing the state variable.

The accuracy and stability of the spatial reconstruction are determined entirely by the choice of the cell slope $(u_s)_i$, with each slope selection defining a distinct numerical scheme.
The simplest choice is to set the slope to zero, $(u_s)_i = 0$, which yields $u_{i+1/2}^- = u_i$ and $u_{i+1/2}^+ = u_{i+1}$.
This corresponds to Godunov's first-order method \cite{Godunov1959}, which is unconditionally non-oscillatory, but introduces heavy numerical dissipation.
A higher-order approximation is obtained using a central difference across neighboring cell centers, $(u_s)_i = (u_{i+1} - u_{i-1})/(2\Delta s)$.  
This simplest non-zero slope yields second-order spatial accuracy in smooth regions:
\begin{equation}
  u_{i+1/2}^- = u_i + \frac{1}{4}(u_{i+1} - u_{i-1})\,, \qquad
  u_{i+1/2}^+ = u_{i+1} - \frac{1}{4}(u_{i+2} - u_i)\,.
\end{equation}
However, by Godunov's theorem \cite{Godunov1959}, any linear scheme with spatial accuracy higher than first order generates unphysical \emph{spurious oscillations} (overshoots and undershoots) near sharp gradients, shocks, or local extrema.

To eliminate unphysical oscillations while retaining second-order accuracy in smooth regions, non-linear \emph{slope limiters} are applied.
A widely used limiter is the generalized \emph{minmod limiter} which compares  central, forward, and backward differences to limit slope magnitude:
\begin{equation}
  (u_s)_i = \operatorname{minmod} \bigg( \theta \frac{u_i - u_{i-1}}{\Delta s}, \,\,
  \frac{u_{i+1} - u_{i-1}}{2\Delta s}, \,\,
  \theta \frac{u_{i+1} - u_i}{\Delta s} \bigg),
  \quad \theta \in [1, 2]\, ,
\label{Eq:minmod_limiter}
\end{equation}
where the multivariable $\operatorname{minmod}$ function is defined as
\begin{equation}
\text{minmod}(a_1, a_2, \dots) = 
\begin{cases} 
  \min(a_1, a_2, \dots) & \text{if } a_j > 0 \;\; \forall j, \\
  \max(a_1, a_2, \dots) & \text{if } a_j < 0 \;\; \forall j, \\
  0 & \text{otherwise.}
\end{cases}
\end{equation}

Near local extrema, forward and backward slopes have opposite signs, causing $\operatorname{minmod}$ to set $(u_s)_i = 0$.
This locally drops the scheme to first order to prevent spurious oscillations.
In monotonic regions, it selects the gentlest slope to preserve \emph{total variation diminishing} (TVD) stability.
For a discrete solution, the total variation is defined as $\displaystyle TV(u) = \sum_i |u_{i+1} - u_i|$.
The TVD scheme guarantees that this quantity never increases during evolution, $TV\big(u(k+\Delta k)\big) \le TV\big(u(k)\big)$, thereby preventing the generation of new, artificial extrema, which would occur in case of oscillations.

Strictly speaking, the standard TVD property applies to space-independent flux $F=F(u)$ in the absence of source terms.
In the present formulation, both the source term $S(s,k)$ and the explicit $s$-dependence of $F$, through the first term of Eq.~\eqref{Eq:dF_chain}, can generate new extrema.
Therefore, the slope limiter guarantees specifically that the spatial reconstruction step itself introduces no artificial numerical oscillations, while allowing physical extrema to evolve under the influence of the source and explicit spatial flux terms.

\subsection{The Kurganov--Tadmor central scheme}

Having constructed the non-oscillatory interface states $u_{i+1/2}^-$ and $u_{i+1/2}^+$ via TVD slope-limited reconstruction, the next step in the finite-volume solver is to evaluate the numerical \emph{interface flux} $\hat{F}_{i+1/2}$.
The Kurganov--Tadmor (KT) scheme \cite{Kurganov:2000ovy} is a high-resolution, semi-discrete central scheme that constructs this numerical flux directly from the reconstructed interface states together with a local estimate of the fastest left- and right-going propagation speeds.
This is achieved without explicitly solving a Riemann problem, as in HLL~\cite{HLL} and HLLC~\cite{HLLC}, or performing a characteristic wave decomposition, as in Roe's method~\cite{Roe}.
At the same time, by adapting the numerical dissipation to the local wave speed rather than relying on a single global bound, the KT scheme avoids the excessive numerical smearing characteristic of global central schemes such as Lax--Friedrichs.

Rather than resolving the internal wave structure of each local Riemann problem, the scheme calculates the local maximum propagation speed at every interface $s_{i+1/2}$:
\begin{equation}
  a_{i+1/2} = \max\!\Big( \Big| F'\big(s_{i+1/2},u_{i+1/2}^-,k\big) \Big|,\ \Big| F'\big(s_{i+1/2},u_{i+1/2}^+,k\big) \Big| \Big),
\end{equation}
where $F' \equiv \partial F/\partial u$, evaluated at fixed $s$ and $k$, is the local characteristic speed (the eigenvalue of the flux Jacobian).
The explicit $s$-dependence of $F$, appearing in the first term of Eq.~\eqref{Eq:dF_chain}, carries no field-space information and thus does not contribute to $a_{i+1/2}$.

The advective KT flux at the interface then combines a central approximation with an adaptive dissipation term into the numerical \emph{advective flux}:
\begin{equation}
  H_{i+1/2} = \frac{1}{2}\Big[F\big(s_{i+1/2},u_{i+1/2}^-\big) + F\big(s_{i+1/2},u_{i+1/2}^+\big)\Big] + \frac{a_{i+1/2}}{2}\big(u_{i+1/2}^+ - u_{i+1/2}^-\big).
  \label{Eq:adv_flux}
\end{equation}
The first term provides second-order central spatial accuracy, while the second term acts as a built-in \emph{numerical diffusion}, scaled strictly by the local jump $\Delta u_{i+1/2} = u_{i+1/2}^+ - u_{i+1/2}^-$.
For the sign of this term see the comment at the end of this section.

For advection--diffusion PDEs with a nonlinear diffusive term $dQ/ds$, the corresponding numerical \emph{diffusive flux} at the interface $s_{i+1/2}$ is denoted by $P_{i+1/2}$ and is discretized following Ref.~\cite{Chertock:2004} as
\begin{equation}
P_{i+1/2} = Q\!\bigg(\frac{u_{i+1}-u_i}{\Delta s}\bigg).
\label{Eq:diff_flux}
\end{equation}
Here, the spatial derivative at the interface is approximated by the unlimited central difference $(u_s)_{i+1/2}=(u_{i+1}-u_i)/\Delta s$, unlike in the reconstruction of the interface values $u_{i+1/2}^{\pm}$, where the TVD-limited slopes $(u_s)_i$ and $(u_s)_{i+1}$ were used.
Because no additional slope limiting is required for the diffusive contribution, the second-order accuracy of the diffusive flux is preserved.
We note that in the general case, when $Q$  also depends explicitly on $s$ and $u$, the discretization introduced in Kurganov and Tadmor~\cite{Kurganov:2000ovy} is required:
\begin{equation}
  P_{i+1/2} = \frac{1}{2}\bigg[ Q\!\bigg(s_i,u_{i+1/2}^-,\, \frac{u_{i+1} - u_i}{\Delta s} \bigg) + Q\!\bigg(s_{i+1}, u_{i+1/2}^+,\, \frac{u_{i+1} - u_i}{\Delta s} \bigg) \bigg].
  \label{Eq:diff_flux_KT}
\end{equation}

Combining the advective and diffusive contributions gives the total conservative interface flux $\hat{F}_{i+1/2} = H_{i+1/2} - P_{i+1/2}$, or explicitly,
\begin{equation}
\begin{split}
  \hat{F}_{i+1/2} &= \frac{1}{2}\Big[F\big(s_{i+1/2},u_{i+1/2}^-\big) + F\big(s_{i+1/2},u_{i+1/2}^+\big)\Big] + \frac{a_{i+1/2}}{2}\left(u_{i+1/2}^+ - u_{i+1/2}^-\right)
  -Q\!\bigg(\frac{u_{i+1}-u_i}{\Delta s}\bigg)\,.
\end{split}
\label{eq:KTflux}
\end{equation}
The PDE~\eqref{Eq:T-flow_ADF} is thereby reduced to a coupled system of ODEs for the cell averages $u_i(k)$,
\begin{equation}
  \frac{d u_i(k)}{dk} = -\frac{\hat{F}_{i+1/2} - \hat{F}_{i-1/2}}{\Delta s} + S(s_i,k)\,,
\end{equation}
which is subsequently integrated in the evolution parameter $k$ using an adaptive method with step-size control, subject to the Courant--Friedrichs--Lewy (CFL) stability condition\cite{CFL}.

We emphasize that the sign of the numerical diffusion term in Eq.~\eqref{Eq:adv_flux} is opposite to that appearing in Eq.~(4.4) of Ref.~\cite{Kurganov:2000ovy}.
The reason is that Eq.~\eqref{Eq:T-flow_ADF} is integrated directly in the decreasing RG scale $k$, whereas the natural time-like variable of the advection--diffusion analogy is the monotonically increasing flow time $t=-\ln(k/\Lambda)$, related to $k$ by $dk/dt=-k$.
Rewriting Eq.~\eqref{Eq:T-flow_ADF} in terms of $t$ turns it into a genuine forward-time conservation law, $\partial_t u + d\tilde F/ds = d\tilde Q/ds + \tilde S$ with $\tilde F=-kF$, $\tilde Q=-kQ$, $\tilde S=-kS$, to which the original Kurganov--Tadmor construction, and hence its dissipation sign, applies unmodified \cite{KoenigsteinSeries}.
Converting the resulting update for the cell averages back to $k$-stepping via $du_i/dt=-k\,du_i/dk$ transforms the standard $-\tilde a_{i+1/2}\Delta u_{i+1/2}/2$ dissipation term into the $+a_{i+1/2}\Delta u_{i+1/2}/2$ appearing in Eq.~\eqref{Eq:adv_flux}, while leaving the diffusive flux $P_{i+1/2}$ of Eq.~\eqref{Eq:diff_flux} unaffected.
The sign therefore compensates for stepping directly in $k$ rather than in $t$, and renders our equations consistent with Ref.~\cite{Kurganov:2000ovy}.

\subsection{CFL condition and adaptive step-size control}
The semi-discrete system is integrated in $k$ using an explicit Runge--Kutta scheme (RK4) with adaptive step-size control (Cash--Karp).
Hyperbolic stability requires that information does not propagate across more than one cell per step.
This is enforced by satisfying the Courant--Friedrichs--Lewy (CFL) condition, which imposes a step-size ceiling based on the maximum cell-interface propagation speed $a_{\max} = \max_i a_{i+1/2}$, given by $\Delta k_{\mathrm{CFL}} = \nu_{\max} \Delta s / a_{\max}$, where we choose $\nu_{\max} = 0.5$.
Recomputed at each step, this bound caps the step chosen by the local truncation error controller: $\Delta k = \min\!\left( \Delta k_{\mathrm{CFL}},\, \Delta k_{\mathrm{error}} \right)$.
The parabolic diffusive term imposes a stricter scaling, $\Delta k \lesssim \Delta s^2 / \big(2\,|\partial Q/\partial u_s|\big)$, which can dominate on fine grids.
Rather than enforcing an explicit diffusive limit, step sizes exceeding this threshold inflate the embedded local truncation error, causing the adaptive controller to reject the step and naturally restrict $\Delta k$ near the parabolic bound.

\subsection{Implementation details of the finite-volume solver}
\label{sec:implementation_details}

This subsection documents the treatment of the two boundaries at $s=0$ and $s=s_{\max}$, the discretization of the fermionic source term, the estimation of the local propagation speed entering the Kurganov--Tadmor flux, and a few further details beyond those discussed above.

\subsubsection{Grid layout and ghost cells}

Following Ref.~\cite{KoenigsteinSeries}, we adopt a node-centered grid, with the physical domain $[0,s_{\max}]$ covered by $N_s$ nodes located at $s_i=i,\Delta s$, $i=0,\dots,N_s-1$, where $\Delta s=s_{\max}/(N_s-1)$.
Thus, the first and last nodes coincide exactly with the physical boundaries, $s_0=0$ and $s_{N_s-1}=s_{\max}$.
For the numerical solution of the flow equation, the grid is extended by two ghost cells on the left, $i=-2,-1$, and three ghost cells on the right, $i=N_s,N_s+1,N_s+2$, with all ghost-cell nodes separated by the same spacing $\Delta s$ as the physical cells.
Three ghost cells are required at the right edge because the polynomial extrapolation used there employs a four-point stencil and is applied cell-by-cell outward (see below).
Figure \ref{fig:grid_ghost} shows a representative slice of this extended grid.

\begin{figure}[!tb]
  \centering
  \includegraphics{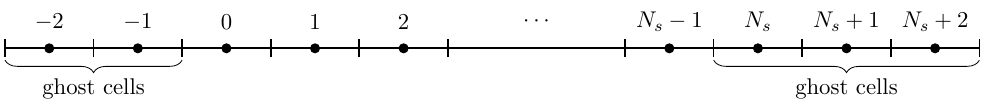}
  \caption{Extended grid used by the solver: two ghost cells at $s_{\min}=0$ ($i=-2,-1$), physical cells $i=0,\dots,N_s-1$ spanning the domain $[0,s_{\max}]$, and three ghost cells at $s_{\max}$ ($i=N_s,N_s+1,N_s+2$).}
  \label{fig:grid_ghost}
\end{figure}

Because $\mathcal{U}_k(\sigma)$ depends on $\sigma$ only through $\rho=\sigma^2/2$, its derivative $u=\partial_\sigma\mathcal{U}_k$ is an odd function of $\sigma$: $u(-s)=-u(s)$.
This property is enforced exactly: after every stage of the evolution in $k$ the cell average at the origin is reset to zero, $u_0=0$, and the two left ghost cells are filled by point reflection through the origin, namely $u_{-1} = -u_{1}$ and $u_{-2} = -u_{2}$.
This reflective boundary condition is what allows the minmod slope at $i=0$ and $i=1$ and the fluxes at the interfaces $s_{\pm1/2}$ adjacent to the origin to be evaluated with the same stencil used in the bulk, without any one-sided formula.

At the outer edge $s=s_{\max}$ of the grid there is no symmetry to exploit.
The three right ghost cells ($i=N_s, N_s+1, N_s+2$) are filled by extending the local cubic polynomial fit determined by the last four physical cells ($u_{N_s-4},\dots,u_{N_s-1}$).
This is implemented sequentially by enforcing a vanishing fourth discrete derivative at each ghost location:
\begin{equation}
u_i = 4 u_{i-1} - 6 u_{i-2} + 4 u_{i-3} - u_{i-4} \quad \text{for } i = N_s, N_s+1, N_s+2\,,
\end{equation}
where each newly computed ghost value is fed forward into the stencil for the next.

\subsubsection{Reduced treatment near the boundaries}

The minmod slope and the dissipative flux are treated specially near the boundaries compared to the interior cells.
The parameter $\theta$ in the minmod limiter \eqref{Eq:minmod_limiter} is normally set to a tunable value $1\le\theta\le2$, but is forced to $\theta=1$ for $i\le1$.
This serves as a safety measure at the coordinate singularity $s=0$: the more compressive choices ($\theta>1$) are only guaranteed non-oscillatory away from the reflective boundary, so the most conservative (and unconditionally TVD) limiter value is used there.
The discretized diffusive term $\big(P_{i+1/2}-P_{i-1/2}\big)/\Delta s$ is set to zero at the boundary cells ($i\le1$ or $i\ge N_s$) rather than evaluated from the interface stencil, preventing ghost values from contaminating the result.
The advective flux $H_{i+1/2}$ and the source term, by contrast, are evaluated uniformly across the entire grid without exception.

\subsubsection{Cell-averaged fermionic source term}

The quark contribution $S(s,k)$ to Eq.~\eqref{Eq:T-flow_ADF} is singular at $T=0$: it contains a Dirac delta at the Fermi surface $E_{\psi,k}(s)=\mu$, where its primitive jumps.
Rather than sampling $S$ pointwise at the cell center, which under- or over-counts the source whenever the Fermi surface falls inside a cell, the solver evaluates the \emph{exact} finite-volume cell average.
This is possible because the fermionic contribution possesses a known analytic primitive $\mathcal{P}(s,k)$ satisfying $\partial_s\mathcal{P} = \pi^2 S/(k^4c_\psi')$, with $c_\psi'=1-\eta_\psi/(d+1)$, yielding
\begin{equation*}
  \langle S\rangle_i = \frac{k^4c_\psi'}{\pi^2} \, \frac{\mathcal{P}(s_{i+1/2},k) - \mathcal{P}(s_{i-1/2},k)}{\Delta s}\,.
\end{equation*}
This exact cell average enters the semi-discrete system, avoiding numerical quadrature errors near the Fermi surface.

\subsubsection{Robust estimate of the local propagation speed}

The propagation speed $a_{i+1/2} = \max(a_1,a_2,a_3,a_4)$ entering the KT flux extends the standard pointwise estimates $a_1=\big|F'(s_{i+1/2},u^-_{i+1/2},k)\big|$ and $a_2=\big|F'(s_{i+1/2},u^+_{i+1/2},k)\big|$ with two additional refinements.
First, the characteristic speed at the average reconstructed state, $a_3 = \big|F'\big(s_{i+1/2},\tfrac12(u^-_{i+1/2}+u^+_{i+1/2}),k\big)\big|$, is included unconditionally.
Second, whenever the reconstructed states straddle zero ($u^-_{i+1/2}\,u^+_{i+1/2}<0$), a secant estimate is incorporated as well:
\begin{equation*}
  a_4 = \frac{\big|F(s_{i+1/2},u^+_{i+1/2},k)-F(s_{i+1/2},u^-_{i+1/2},k)\big|}
             {\max\big(|u^+_{i+1/2}-u^-_{i+1/2}|,\,\epsilon\big)}\,,
\end{equation*}
where $\epsilon = 10^{-10}$ prevents division by zero (setting $a_4=0$ when the states do not straddle zero).
Including these estimates guards against underestimating the wave speed when the flux Jacobian $F'$ varies rapidly between the two reconstructed states, a situation that occurs precisely where $u$ changes sign, near nontrivial extrema of $\mathcal{U}_k$.

\subsubsection{Error control and step-size safeguards}

The scale against which the local truncation error is measured is $\max(|u_i|,\varepsilon_{\mathrm{abs}})+|\Delta k\,du_i/dk|$ rather than the bare $|u_i|+|\Delta k\,du_i/dk|$ of the textbook embedded Runge--Kutta scheme.
The absolute floor $\varepsilon_{\mathrm{abs}}$ prevents the relative error criterion from becoming unachievably strict near the grid points where $u\to0$.
Note that besides $u_0=0$, enforced exactly by the reflective boundary condition, $u$ also flattens to zero over a wide grid region, as the convexity of the potential is restored in the infrared.

Additionally, if adaptive step reduction shrinks $\Delta k$ to the point where $k+\Delta k$ no longer differs from $k$ in floating-point arithmetic, the step is force-accepted to avoid infinite loops.
This typically happens at a minmod kink coincident with a zero crossing of $u$.

\section{Parametrization}\label{sec:param}

The parameters of the ultraviolet potential and the couplings, $m^2$, $\lambda$, $g$, and $h$, are fixed by conditions imposed at the end of the flow, at $k_{\rm IR}$.
Therefore, a parameter set belongs to a given pair $(\Lambda,k_{\rm IR})$ and to a given truncation and, in principle, cannot be carried over to a different cutoff, a different infrared scale, or a different choice of expansion point used in the flows of the $Z$ factors in the LPA$'$ case.
We impose three conditions exactly: the renormalized minimum, $\bar\phi_R=Z_{\pi,k_{\rm IR}}^{1/2}\bar\phi_{k_{\rm IR}}=f_\pi$; the pion mass, through $h=Z_{\pi,k_{\rm IR}}^{1/2}f_\pi m_\pi^2$; and the constituent quark mass, $g\bar\phi_{k_{\rm IR}}/Z_{\psi,k_{\rm IR}}=M_q$.
A fourth condition, on the sigma curvature mass, $m_\sigma=(M^2_{\sigma,k_{\rm IR}}/Z_{\sigma,k_{\rm IR}})^{1/2}$, serves only as a target and is not required to be reproduced exactly.

The three exact conditions leave a one-parameter family of solutions, with $\lambda$ providing a natural coordinate.
At fixed $\lambda$, the remaining three parameters are determined by a Newton-type iteration, and $m_\sigma$ is then a function of $\lambda$ alone.
This function is very flat: the flow largely forgets the ultraviolet value of the quartic coupling, and $m_\sigma$ changes by less than 1~MeV per unit change in $\lambda$ near its extremum.
In LPA$'$, the target value $m_\sigma=500$~MeV is usually not attainable.
In particular, with the running minimum at $\Lambda=500$~MeV, $m_\sigma(\lambda)$ has a maximum of $473.8$~MeV.
This value is therefore a prediction of the truncation rather than an input, whereas in the LPA the target value can be reached.
It proved useful in LPA$'$ to fix $\lambda$ and impose only the three exact conditions.
With the fixed expansion point $\phi_\star=96$~MeV, this gives $m_\sigma=500.4$~MeV at $\lambda=11.46$ and converges in a few iterations, whereas searching for $\lambda$ is computationally expensive and, far from the solution, leads to very stiff flows.

\begin{table}[!t]
\centering
\begin{tabular}{ccc||cccc||cccc}
\hline
 & $\phi_\star$ [MeV] & $\Lambda$ [MeV] & $m^2$ [MeV$^2$] & $\lambda$ & $g$ & $h$ [MeV$^3$] & $m_\sigma$ [MeV] & $Z_\pi$ & $Z_\sigma$ & $Z_\psi$\\
\hline
LPA    & N/A                  & 500 & $2.221521$E3  & $10.853702$ & $3.365591$ & $1.771092$E6 & $500.00$ & $1$      & $1$      & $1$\\
LPA & N/A & 500 & $2.972388$E4 & $8.259446$ & $3.365591$ & $1.771092$E6 & $473.82$ & $1$ & $1$ & $1$\\
LPA$'$ & $96$                 & 500 & $-5.072430$E3 & $11.45833$  & $3.603443$ & $1.755251$E6 & $500.44$ & $0.9822$ & $1.0551$ & $1.0803$\\
LPA$'$ & $\bar\phi|_{\rm g}$   & 500 & $-1.307674$E2 & $14.061387$ & $3.935982$ & $1.862638$E6 & $473.82$ & $1.1061$ & $1.3166$ & $1.1120$\\
LPA$'$ & $96$                 & 1000 & $9.998889$E5 & $12$        & $4.459353$ & $2.159628$E6 & $690.25$ & $1.4869$ & $1.0117$ & $1.0866$\\
\hline
\multicolumn{3}{c||}{} & \multicolumn{4}{c||}{UV parameters} & \multicolumn{4}{c}{IR values}\\
\end{tabular}
\caption{Parameters of the three truncations, obtained at $k_{\rm IR}=40$~MeV ($m\mathrm{E}n$ corresponds to $m\times 10^{n}$).
  $\phi_\star$ is the expansion point of the $Z$ flows in LPA$'$, which is either fixed or taken to be the running global minimum $\bar\phi|_{\rm g}$ of the effective potential.
  The UV parameters are fitted to the target values $\bar\phi_R=93$~MeV, $M_q=313$~MeV, $m_\pi=138$~MeV and $m_\sigma=500$~MeV at $T=\mu=0$.
  The first three are attained exactly.
  As for $m_\sigma$, at $\Lambda=500$~MeV it is attained in the LPA and comes out close to the target value in the case of LPA$'$ with $\phi_\star$ fixed, while with the running minimum $473.82$~MeV is the largest value that can be reached.
  The second LPA set is fitted with this mass value in order to compare the location of the critical endpoints at the same vacuum sigma mass.
  At $\Lambda=1000$~MeV, with $\phi_\star$ fixed and $\lambda=12$ kept fixed during the fit, $m_\sigma$ comes out well above the target value.
}
\label{tab:params}
\end{table}

\begin{figure}[!t]
  \centering
  \includegraphics[width=0.5\textwidth]{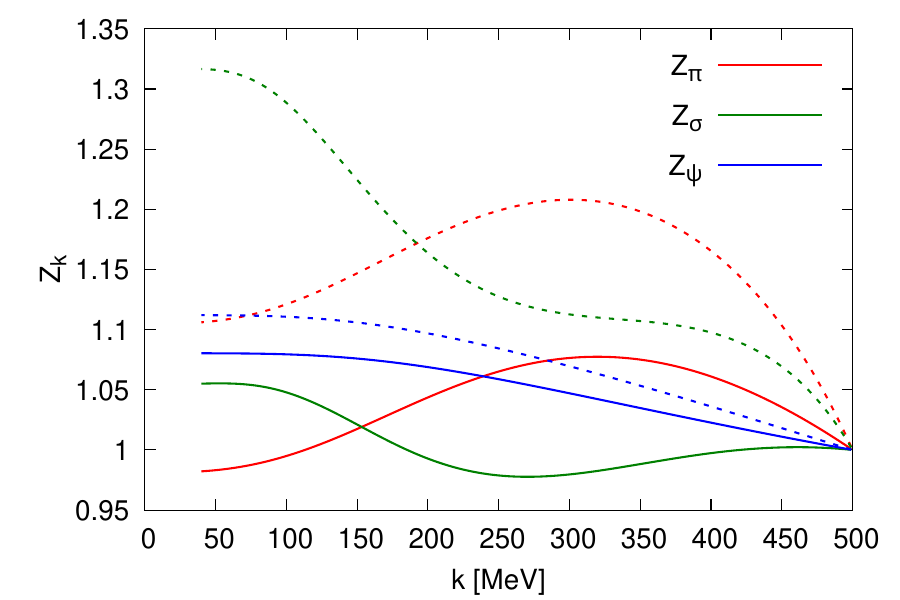}
  \caption{The wave-function renormalizations $Z_{\pi,k}$, $Z_{\sigma,k}$, and $Z_{\psi,k}$ as functions of the RG scale $k$ in the vacuum ($T = \mu = 0$) for UV scale $\Lambda = 500~\text{MeV}$ and $k_{\text{IR}} = 40~\text{MeV}$. 
    The $Z$ factors evaluated at the fixed expansion point $\phi_\star = 96~\text{MeV}$ are plotted as solid lines, while those evaluated at the running minimum of the potential are plotted as dashed lines. 
    Each set of curves is calculated using its corresponding parameter set listed in Table~\ref{tab:params}.}
  \label{fig:Z_vac}
\end{figure}

The parameters used in the three truncations investigated in this work are listed in Table~\ref{tab:params}.
They were determined in three stages, in which the flow is always integrated from $\Lambda$, but the conditions are imposed at successively lower scales, $k_{\rm IR}=100$, $60$ and finally $40$~MeV.
The first stage, where the flows are computationally inexpensive, explores the family of solutions: at each value of $\lambda$ visited it solves the three exact conditions and records $m_\sigma$, and so maps $m_\sigma(\lambda)$ and locates the $\lambda$ at which $m_\sigma$ is closest to the target.
The later stages follow the solution down to the final $k_{\rm IR}$, each starting from the solution obtained at the previous stage, with $\lambda$ either refined in small steps or kept fixed.
The starting point of the first stage is obtained by a cold start, which needs no prior knowledge of the parameters.
Setting the $Z$ factors to one, the targets give first estimates $h\simeq f_\pi m_\pi^2$ and $g\simeq M_q/f_\pi$, and $m^2$ is then determined as the zero of $\bar\phi_R-f_\pi$, found by a bracketing method.
For this step the flow is integrated only down to $k\simeq\Lambda/3$, where it is inexpensive.
The $Z$ factors obtained there are used to correct $g$ and $h$, and the determination of $m^2$ is repeated a few times before the first stage starts.
The resulting $k$ dependence of the $Z$ factors in the vacuum is shown in Fig.~\ref{fig:Z_vac} for the two LPA$'$ parametrizations of Table~\ref{tab:params}.

In the LPA$'$ truncation with the current version of the flow equations, we could parameterize the model at $\Lambda = 1~\text{GeV}$ only with the $Z$ factors evaluated at a fixed expansion point.
Using $\phi_\star = 96~\text{MeV}$, where the quarks are heavy throughout the flow, the vacuum observables $f_\pi$, $m_\pi$, and $M_q$ are matched precisely.
However, the sigma mass $m_\sigma$ exceeds the target value, yielding values between $680$ and $700$~MeV for $\lambda$ between $4$ and $20$.
When the $Z$ factors are instead evaluated at the running minimum of the potential, no viable parameter set could be found.
For a symmetric UV potential, the minimum stays close to the origin over a large portion of the flow, where the quarks are nearly massless and the quark loop induces a large pion anomalous dimension $\eta_\pi$.
Increasing the Yukawa coupling $g$, which is required to reach the target constituent quark mass $M_q$, further enhances $Z_\pi$.
As a result, $M_q$ grows only very slowly with $g$ and does not exceed about $210~\text{MeV}$ before the flow breaks down, even if target values for $M_q$ between $300$ and $400~\text{MeV}$ are accepted.

We could not find a set of parameters that reproduces the $k$-dependence of $Z_{\pi,k}$ and $Z_{\sigma,k}$ reported for $\Lambda=1~\text{GeV}$ in Fig.~11 of Ref.~\cite{Weyrich:2018jid}.
This makes a direct comparison with those results impossible.

The parametrization discussed here was done with a code using the standard grid method on a uniform $\rho$-grid.
We chose $\Delta\rho\simeq14.43$~MeV$^2$, while the largest value of the field, and therefore the number of grid points, was fixed by $\Lambda$ through $\phi_\textrm{max}/\Lambda=0.24$ (i.e.\ $\rho_\textrm{max}=7200$~MeV$^2$ and $500$ points at $\Lambda=500$~MeV, $\rho_\textrm{max}=28800$~MeV$^2$ and $1997$ points at $\Lambda=1$~GeV).

\section{LPA results}\label{sec:LPA}

In this section, we compare the LPA results obtained using the KT scheme with those derived from the standard grid method.
We first consider the chiral limit before turning to the case of a physical pion mass.

\subsection{Chiral limit}

We adopt the UV parameters of Ref.~\cite{Schaefer:2004en} and do the comparison at $T=0.01\text{~MeV}$ and $\mu=241\text{~MeV}$, where our results reproduce the findings reported in that reference.
In the grid method, we solve the flow equation for the effective potential, whereas in the KT scheme, we solve the flow equation for $u(\sigma)$, the first derivative of the effective potential with respect to the scalar background field.

The KT scheme is implemented in Fortran using a $\sigma$-grid as described in the previous section, whereas the standard grid method is implemented in C, with the potential discretized on a uniform $\rho$-grid.
In both cases, the grid contains $N_s=500$ number of points (or cells), covering the field domain up to a maximum value of $\sigma_{\max}=120\text{~MeV}$.
In the grid method, the first and second derivatives of the potential, which are required during the flow for the evaluation of the pion and sigma curvature masses, are calculated using second-order central finite difference formulas, except at the two boundary points, where forward and backward approximations of the same order are used.
For adaptive step-size ODE integration, we employ the explicit embedded Runge–Kutta–Fehlberg (4,5) method from the GNU Scientific Library with an accuracy requirement of $10^{-14}$.
This high accuracy is mainly required to monitor the flattening of the effective potential in the deep infrared.

\begin{figure}[!tb]
  \centering
  \includegraphics[width=0.9\textwidth]{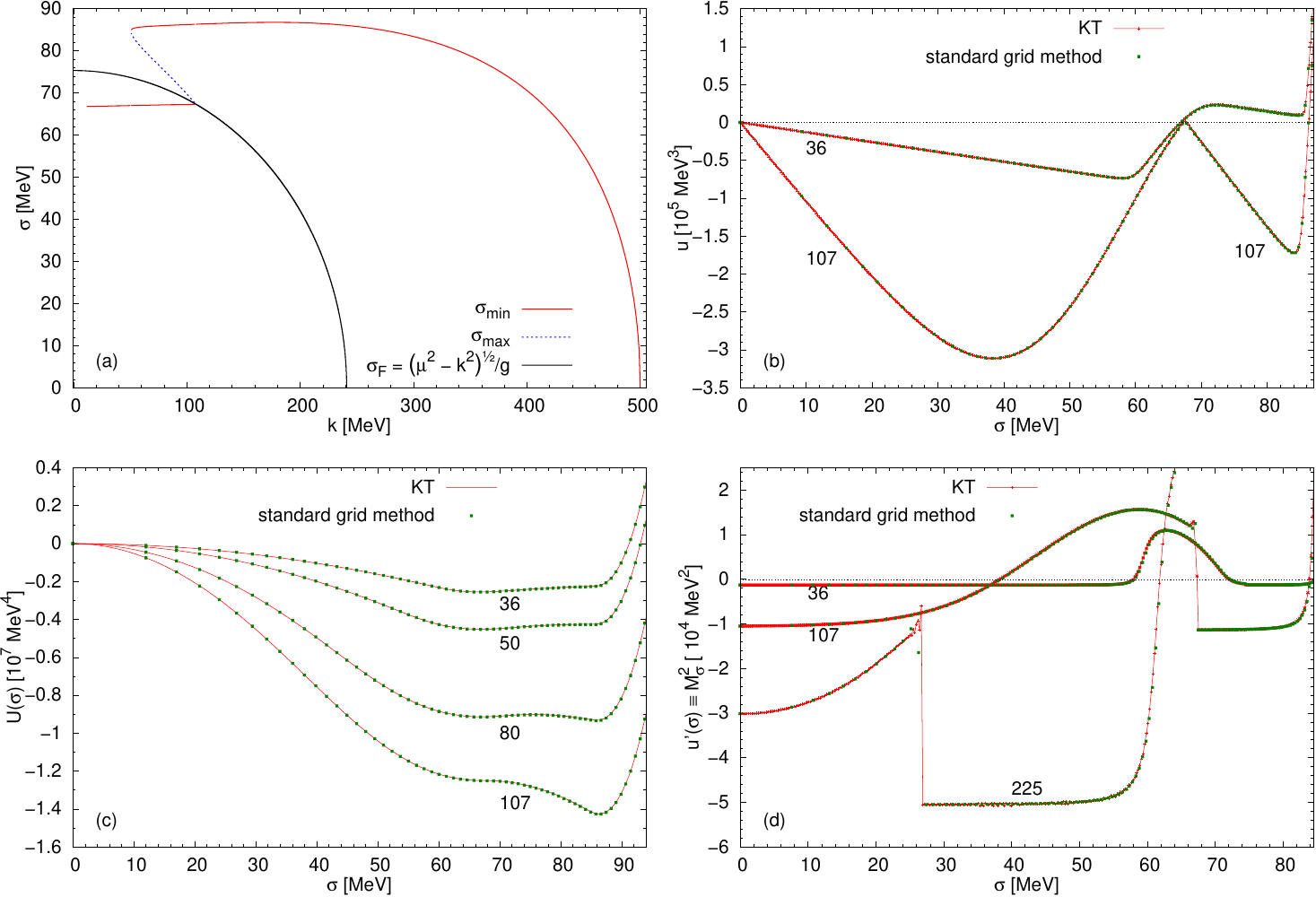}
  \caption{Comparison of results from the KT scheme and the standard grid method, obtained using $N_s=500$ cells and grid points, respectively, at $T=0.01$~MeV and $\mu=241$~MeV with the UV parameters of Ref.~\cite{Schaefer:2004en}.
    Panel (a) shows the $k$-dependent extremum of the effective potential relative to the Fermi edge (surface).
    Panel (c) displays the effective potential at various values of $k$, while panels (b) and (d) show its first and second derivatives.
    The numerical labels indicate the value of $k$ in MeV.}
  \label{fig:lpa_comparison}
\end{figure}

As is well known from mean-field analysis, the first derivative of the effective potential is continuous at the Fermi surface $\sigma_F$, while the second derivative is discontinuous.
Consequently, the sigma curvature mass inside the Fermi surface (i.e. for $\sigma<\sigma_F$) differs from its value outside.
These features are clearly captured in panels (b) and (d) of Fig.~\ref{fig:lpa_comparison}.
The Fermi surface itself is shown in panel (a) alongside the $k$-dependence of the nontrivial extrema of the effective potential, that is the values of $\sigma$ where $u(\sigma) = 0$.
The jump in $u'(\sigma)$ is largest for $k$ close to the Fermi momentum; its magnitude gradually decreases towards the infrared, becoming barely visible at $k=36$~MeV.
Panel (c) shows the evolution of the effective potential with the scale parameter $k$ as it settles into an IR profile, demonstrating that the chosen chemical potential $\mu=241$~MeV is smaller than the critical value $\mu_c$ at which the potential has two degenerate nontrivial minima: a broken phase minimum and a dense phase minimum.

The two methods yield virtually identical results across all scale parameters.
This excellent agreement confirms that, at this grid resolution, the standard grid method handles the non-analyticities near the Fermi surface with high fidelity, validating its accuracy in this regime.

\subsection{Physical pion mass \label{ss:lpa_phys}}

In the case of the physical pion mass, we first verified that both the KT scheme and the standard grid method reproduce the IR quantities obtained with the UV parameters reported in the literature.
In particular, we reproduced the parameters used in Ref.~\cite{Tripolt:2013jra, Tripolt:2017zgc} and the parameters obtained at $\Lambda=1$~GeV in Ref.~\cite{Stoll:2026ezu}.
Since these parameters are obtained for $\Lambda>500$~MeV, we parameterized the model at $\Lambda=500$~MeV, the value used in the chiral limit.
To this end, we used the procedure described in Sec.~\ref{sec:param}.
The parameter values used in this section are listed in Table~\ref{tab:params}.

\begin{figure}[!tb]
  \centering
  \includegraphics[width=0.495\textwidth]{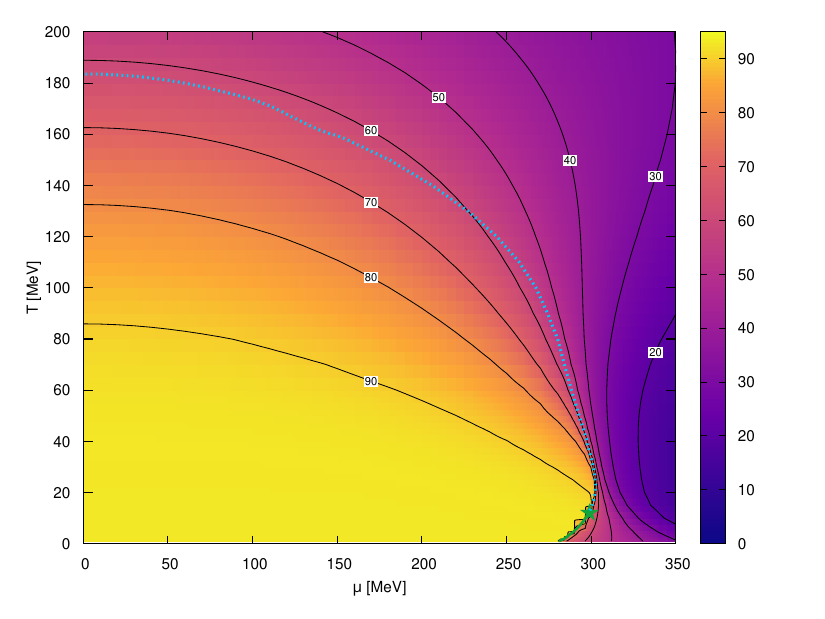}\hfill
  \includegraphics[width=0.495\textwidth]{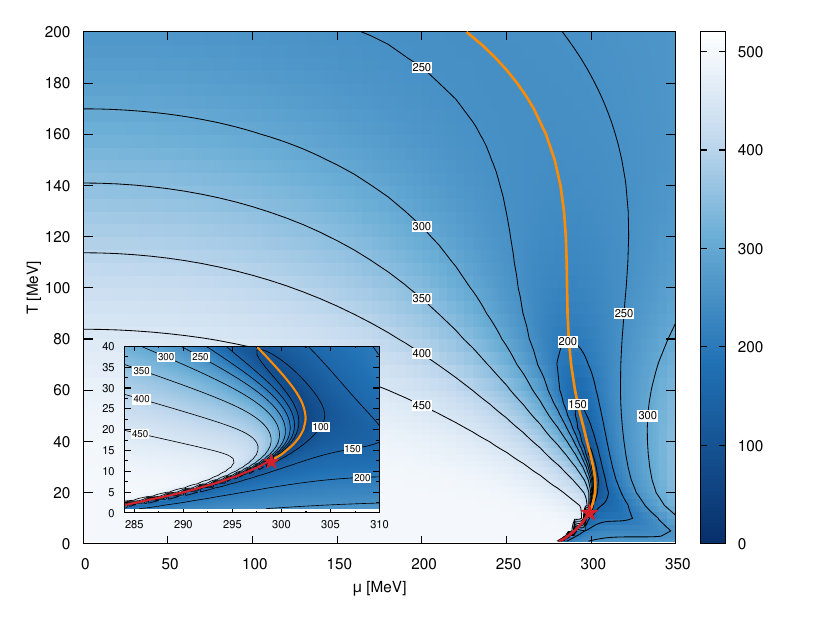}
  \caption{Phase diagram of the QM model in the LPA ($\Lambda=500$~MeV, $k_{\rm IR}=40$~MeV), with the parameters fixed using the physical pion mass and $m_\sigma=500$~MeV.
    Left panel: color map and contour lines of the order parameter $\bar\phi$ (in MeV).
    The cyan dotted line is the ridge of $|\nabla\bar\phi|=[(\partial_\mu\bar\phi)^2+(\partial_T\bar\phi)^2]^{1/2}$, where $\bar\phi$ changes fastest: the maximum of $|\nabla\bar\phi|$ in $\mu$ at fixed $T$, and, close to the $T$ axis, in $T$ at fixed $\mu$.
    Right panel: color map and contour lines of the $\sigma$ curvature mass $M_\sigma$ (in MeV).
    The orange line is the ridge of the chiral susceptibility $\chi=\partial\bar\phi/\partial h=1/M_\sigma^2$, the maximum of $\chi$ in $\mu$ at fixed $T$.
    The inset zooms on the region around the CEP.
    In both panels the star marks the estimated location of the CEP (see Table~\ref{tab:CEP} for coordinates), and the first-order line is drawn in green (left) and red (right).}
  \label{fig:LPA_k40}
\end{figure}

Since our implementation of the KT scheme turned out to be computationally far more costly than the standard grid method, we used the latter to scan the $T$--$\mu$ plane, both to determine the phase boundary and to obtain a reference against which the LPA$'$ results can be compared.

The phase diagram of the model is shown in Fig.~\ref{fig:LPA_k40}.
The crossover transition is characterized in two ways: by the ridge of $|\nabla\bar\phi|$, along which the order parameter changes fastest in the $T$--$\mu$ plane, and by the ridge of the chiral susceptibility $\chi=\partial\bar\phi/\partial h=1/M_\sigma^2$, i.e.\ the line of minimal $M_\sigma$ at fixed $T$.
The two lines are close to each other near the CEP, where the crossover is sharp, and separate at higher temperatures, where it becomes broad.
The first-order transition line ends at the CEP with coordinates given in Table~\ref{tab:CEP}, and, as expected from the discussion in the Introduction for the Litim regulator, it exhibits the characteristic backbending.

We also determine the CEP in LPA for the vacuum mass $m_\sigma=473.82$~MeV obtained in the parametrization of LPA$'$ with the running minimum (second row of Table~\ref{tab:params}).
By bisection in $\mu$ and examination of the potential at $k_{\rm IR}$, we find a first-order transition at $T=16$~MeV (two coexisting minima at $\mu=296.514$~MeV) and a crossover at $T=16.5$~MeV ($M_\sigma$ reaches only $5.6$~MeV at $\mu_{pc}=296.664$~MeV), hence the CEP lies between these two temperatures (see Table~\ref{tab:CEP}).

\begin{table}[!t]
\centering
\begin{tabular}{cccc||cc}
\hline
  & $\phi_\star$ [MeV] & $\Lambda$ [MeV] & $m_\sigma$ [MeV] & $T_{\rm CEP}$ [MeV] & $\mu_{\rm CEP}$ [MeV]\\
\hline
LPA & N/A & 500 & $500.00$ & $\simeq12.5$ & $\simeq299$\\
LPA & N/A & 500 & $473.82$ & $16$--$16.5$ & $\simeq296.6$\\
LPA$'$ & $96$ & 500 & $500.44$ & $7$--$7.5$ & $\simeq296$\\
LPA$'$ & $\bar\phi|_{\rm g}$ & 500 & $473.82$ & $8.75$--$9.2$ & $\simeq298.8$\\
LPA$'$ & $96$ & 1000 & $690.25$ & $7.0$--$7.2$ & $\simeq296$\\
\hline
\end{tabular}
\caption{Location of the critical endpoint for the parameter sets of Table~\ref{tab:params} ($k_{\rm IR}=40$~MeV).}
\label{tab:CEP}
\end{table}

\section{$\mathrm{LPA}'$ results}\label{sec:LPAp}

Within the $\mathrm{LPA}'$, we investigate the challenges arising from the inclusion of the $Z$ factors.
First, we present the flow equations for the wave function renormalization factors, derived from their standard definitions given in Eqs.~\eqref{Eq:Zb_def} and \eqref{Eq:Zf_def} using the method applied also in Ref.~\cite{Fejos:2025qlv}.
We then present the numerical results obtained within the $\mathrm{LPA}'$ approximation and discuss how the flowing $Z$ factors, through the anomalous dimensions, affect the flow of the effective potential and the resulting phase diagram.

\subsection{Flow of the wave function renormalization factors}\label{ss:Zflow}

The bosonic contribution to the flow of $Z_{\pi,k}$ and $Z_{\sigma,k}$ was calculated in Ref.~\cite{Weyrich:2018jid} by neglecting $Y_k$ in the vertices, that is, by working with momentum-independent vertices.
We use the same approximation in three-point vertices, but include the contribution of $Y_k$ coming from the tadpole integral.
At finite temperature and in $d$ spatial dimensions we obtain from the formulas given in Appendix~\ref{App:2point} and the definition \eqref{Eq:Zb_def}
\bea
\partial_k Z_{\pi,k}&=&-2\frac{\Omega_d}{d}\,\frac{k^{d+1}}{Z_{\pi,k} Z_{\sigma,k}}\,\big(\phi_\star\, U_k''(\rho_\star)\big)^2\,
\frac{\partial^2 B(E_{\pi,k},E_{\sigma,k})}{\partial E^2_{\pi,k}\,\partial E^2_{\sigma,k}}
-\frac{\Omega_d}{d} k^{d+1}\frac{Y_k}{Z_{\pi,k}} B(E_{\pi,k}) \bigg(1-\frac{\eta_{\pi}}{d+2}\bigg)
\nonumber\\
&&
-4\frac{\Omega_d}{d}\frac{k^{d-1}}{(d-2)Z^2_{\psi,k}}
g^2 N_c
\left[
\big(2\eta_\psi-d\big) F_2(E_{\psi,k}) - 4 k^2 \big(\eta_\psi-(d-1)\big) F_3(E_{\psi,k})
\right] \,,
\label{Eq:Zpi_flow}
\\
\partial_k Z_{\sigma,k}&=&-(N-1)\frac{\Omega_d}{d}\frac{k^{d+1}}{Z^2_{\pi,k}}
\big(\phi_\star U_k''(\rho_\star)\big)^2
\, \bigg[\frac{1}{6} \frac{d^2 B(E_{\pi,k})}{dE^2_{\pi,k} dE^2_{\pi,k}}\bigg]
\nonumber\\
&&-\frac{\Omega_d}{d}\frac{k^{d+1}}{Z^2_{\sigma,k}}\big(3\phi_\star U_k''(\rho_\star)+\phi_\star^3 U_k^{(3)}(\rho_\star)\big)^2\,
\bigg[\frac{1}{6}\frac{d^2 B(E_{\sigma,k})}{dE^2_{\sigma,k} dE^2_{\sigma,k}}\bigg]
-\frac{\Omega_d}{d} k^{d+1}\frac{Y_k}{Z_{\sigma,k}} B(E_{\sigma,k}) \bigg(1-\frac{\eta_{\sigma}}{d+2}\bigg)
\nonumber\\
&&-4\frac{\Omega_d}{d}\frac{k^{d-1}}{(d-2)Z^2_{\psi,k}} g^2 N_c
\Big[
  \big(2\eta_\psi-d\big) F_2(E_{\psi,k}) - 4 k^2\big(\eta_\psi-(d-1)\big) F_3(E_{\psi,k}) \nonumber\\
  &&\hspace*{4.25cm}
  - 8(d-2)k^2\big(E_{\psi,k}^2-k^2\big) F_4(E_{\psi,k})
  \Big]\,,
\label{Eq:Zsigma_flow}
\eea
where the terms proportional to $Y_k$ represent tadpole contributions and in $d=3$, one has $\Omega_3/3=1/(6\pi^2)$.
The factor of two in the bosonic contribution to $\partial_k Z_{\pi,k}$ accounts for the fact that, in the bubble integral containing the product of propagators, $G^R_{\pi,k} G^R_{\sigma,k}$, differentiating either $G^R_{\pi,k}$ or $G^R_{\sigma,k}$ with respect to the regulator gives the same contribution.

The functions appearing in the bosonic contributions in Eqs.~\eqref{Eq:Zpi_flow} and \eqref{Eq:Zsigma_flow} are defined as
\bea
B(E_1,E_2) &=& T\sum_m\frac{1}{(\omega_m^2 + E_1^2)(\omega_m^2 + E_2^2)}
=
\frac{E_1 \left(1 + 2n(E_2)\right) - E_2 \left(1 + 2n(E_1)\right)}{2 E_1 E_2 \left(E_1^2 - E_2^2\right)},\
\label{Eq:B12_def}
\\
B(E_1)&=&
\lim_{E_2\to E_1} B(E_1,E_2)=
\frac{1 + 2n(E_1) - 2E_1 n'(E_1)}{4E_1^3},\ 
\label{Eq:B1_def}
\eea
where $\omega_m = 2\pi n T$ are the bosonic Matsubara frequencies and $n(E)=1/(e^{E/T}-1)$ is the Bose--Einstein distribution function.
The functions $F_n(E)$, $n\ge 2$, appearing in the fermionic contribution are defined as in Ref.~\cite{Pawlowski:2014zaa} through the recursion $\displaystyle F_n=-\frac{1}{n-1}\frac{d F_{n-1}}{d E^2}$,
with
\be
F_1(E)=T\sum_n\frac{1}{(\nu_n\pm \ri\mu)^2 + E^2} = \frac{1}{2 E} \big[1-f_+(E)-f_-(E)\big]\,.
\label{Eq:F1}
\ee
Here $\nu_n = (2n+1)\pi T$ are the fermionic Matsubara frequencies and $f_\pm(E)=1/(e^{(E\mp \mu)/T}+1)$ are the Fermi--Dirac distribution functions for particles and antiparticles.

Using the explicit expressions for the derivatives of $B(E_1,E_2)$, $B(E_1)$ and $F_1(E)$, which we do not give here, one can check that Eqs.~\eqref{Eq:Zpi_flow} and \eqref{Eq:Zsigma_flow} agree with Eqs.~(C.22)--(C.26) of Ref.~\cite{Weyrich:2018jid}, when the tadpole contributions are neglected in the former and after taking into account that $\rho$ in Ref.~\cite{Weyrich:2018jid} is twice ours\footnote{There is, however, a typo in Eq.~(C.26) of Ref.~\cite{Weyrich:2018jid}: $\sigma_0$ should appear squared.}.

The flow of the fermionic wave function renormalization factor, obtained with the projection \eqref{Eq:Zf_def}, is
\bea
\partial_k Z_{\psi,k} &=& 2 \frac{\Omega_d}{d}  k^{d+1} g^2 \bigg[
  (N-1)Z^{-1}_{\pi,k}Z^{-1}_{\psi,k}\Big(1-\frac{\eta_{\pi}}{d+1}\Big) \mathfrak{R}\frac{ \partial B_{fb}(E_{\psi,k},E_{\pi,k};P^0_\textrm{low})}{\partial E^2_{\pi,k}} \nonumber \\
  && \hspace*{2.5cm} + Z^{-1}_{\sigma,k} Z^{-1}_{\psi,k} \Big(1-\frac{\eta_{\sigma}}{d+1}\Big) \mathfrak{R}\frac{ \partial B_{fb}(E_{\psi,k},E_{\sigma,k};P^0_\textrm{low})}{\partial E^2_{\sigma,k}} \bigg]\,,
\label{Eq:Zpsi_flow}
\eea
where the external fermionic frequency is set to $P^0_\textrm{low}=\ri\pi T-\mu$, as dictated by Eq.~\eqref{Eq:Zf_def}.
For $d=3$ and a constant Yukawa coupling this agrees with Eq.~(3.21) of Ref.~\cite{Pawlowski:2014zaa}, once we take into account the relation \eqref{Eq:eta_def} between $\eta_\psi$ and $Z_{\psi,k}$ and the fact that the derivative corresponds to the function $\mathcal{FB}_{(1,2)}$ introduced there.
This derivative is defined in terms of the Matsubara sum given in Ref.~\cite{Bellac:2011kqa} for a fermionic external frequency $P_0=\ri \nu_m - \mu$ and a loop frequency variable $Q_0=\ri \nu_n - \mu$, with $\nu_m=(2m+1)\pi T$, for which the difference $Q_0-P_0=\ri 2(n-m)\pi T=:\ri\omega_{n-m}$ is then bosonic:
\bea
B_{fb}(E_f,E_b;P_0) &=& T\sum_n \frac{1}{(E_f^2-(\ri \nu_n - \mu)^2)(E_b^2-(Q_0-P_0)^2)}\nonumber\\
&=&-\sum_{s_f,s_b=\pm}\frac{s_f s_b}{4E_f E_b} \frac{1-f_-(s_f E_f) + n(s_b E_b)}{P_0-s_f E_f-s_b E_b}\,,
\label{Eq:BF_Msum}
\eea
where the subscripts $f$ and $b$ refer to fermionic and bosonic quantities.
The derivative of Eq.~\eqref{Eq:BF_Msum} that appears in Eq.~\eqref{Eq:Zpsi_flow} is given compactly by
\begin{align}
\!\!\!\mathfrak{R}\frac{\partial B_{fb}(E_f,E_b;P^0_\textrm{low})}{\partial E_b^2} = &
\sum_{s=\pm}\left[\frac{8\pi^2T^2\,(E_f+s\mu)^2}{(\Delta^D_s)^2\,(\Delta^S_s)^2}
-\frac{1}{\Delta^D_s\,\Delta^S_s}\right]\frac{1-2f_{-s}(E_f)}{4E_f}
\nonumber\\[2pt]
&+\sum_{s=\pm}\left[\frac{1}{\Delta^D_s}-\frac{1}{\Delta^S_s}
+2\pi^2T^2\left(\frac{1}{(\Delta^S_s)^2}-\frac{1}{(\Delta^D_s)^2}\right)\right]
\frac{1+2n(E_b)}{16E_fE_b^2}
\nonumber\\[2pt]
&-\sum_{s=\pm}\left[\frac{S_s}{\Delta^S_s}+\frac{D_s}{\Delta^D_s}\right]
\frac{1+2n(E_b)-2E_b\,n'(E_b)}{16E_fE_b^3},
\label{Eq:B12_P0low}
\end{align}
\noindent where $S_s = E_f + E_b + s\mu$, $D_s = E_f - E_b + s\mu$, $\Delta^X_s = \pi^2T^2 + X_s^2$ for $X=S,D$, and $s=\pm$.
Evaluating this derivative instead at vanishing external frequency, $P^0=0$, would give an expression that is singular at $E_f=E_b$.

Using $P^0_\textrm{low}$, as we did above, leads to an unphysical $\mu$ and $T$ dependence in Eq.~\eqref{Eq:B12_P0low}, since these do not enter only through the Fermi--Dirac statistical factors, as they would in the real-time formalism of finite-temperature QFT.
The extra $\mu$-dependence leads to a violation of the Silver Blaze property occurring at $T=0$, which was described in QCD in Ref.~\cite{Cohen:2003kd} and for a complex scalar model treated within the 2PI formalism in Ref.~\cite{Marko:2014hea}.
The remedy is known in the FRG literature \cite{Fu:2015naa, Fu:2019hdw}.
The prescription of Ref.~\cite{Fu:2019hdw} consists of replacing $P^0_\textrm{low}$ by $P^0_\textrm{ex}=\ri y_k$, with $y_k=\pi T \exp\{-k/(\pi T)\}$.
The resulting expression for $\mathfrak{R}\frac{\partial B_{fb}(E_f,E_b;P^0_\textrm{ex})}{\partial E_b^2}$ can be obtained from that in Eq.~\eqref{Eq:B12_P0low} by neglecting $\mu$ outside the statistical factors and replacing $\pi T$ with $y_k$.
The sum over $s$ in the fermionic contribution is then proportional to $F_1(E_f)$, where $F_1$ is defined in Eq.~\eqref{Eq:F1}.
However, using $P^0_\textrm{ex}$ brings back the singularity problem mentioned below Eq.~\eqref{Eq:B12_P0low}: with this modification the derivative has a double pole at $E_f=E_b$, regulated only by $y_k$, which is exponentially small for $k\gg\pi T$.
We indeed met this pole at $T=0$ when the expansion point lies inside the Fermi sea, where Pauli blocking removes the term that cancels it in the vacuum.
To avoid this problem, which stops the integration of the flow, we work with Eq.~\eqref{Eq:B12_P0low} when exploring the $T$--$\mu$ plane.
We checked that at low temperatures the first-order line obtained with $P^0_\textrm{ex}$ still bends back and lies only slightly higher in $\mu$ than with Eq.~\eqref{Eq:B12_P0low}.


The equations \eqref{Eq:Zpi_flow}, \eqref{Eq:Zsigma_flow}, and \eqref{Eq:Zpsi_flow} were implemented in both codes used for the LPA: the one based on the KT scheme, which evolves the field derivative of the effective potential, and the one based on the standard grid method for the effective potential itself.
In the grid method, the three $Z$ factors are added as further components to the system of ordinary differential equations for the values of the potential at the grid points and are integrated simultaneously with it, the anomalous dimensions being obtained at each evaluation of the right-hand side by solving the linear system at the expansion point.
In the KT scheme, the two flows are integrated separately: in each RG step, the field derivative of the potential is advanced with the $Z$ factors and anomalous dimensions kept at their values at the beginning of the step, after which the $Z$ factors are integrated over the same step, in a few substeps, using the updated potential.
As in LPA, the grid method was used to produce the data shown in the figures, and the KT scheme to cross-check selected points in the $T$--$\mu$ plane.

The flows of $Z_{\pi,k}$, $Z_{\sigma,k}$, and $Z_{\psi,k}$ require $U_k'$, $U_k''$, and $U_k^{(3)}$ at the expansion point at every evaluation of the right-hand side of the system of ordinary differential equations.
Care has to be taken in the way these derivatives are evaluated, as an incorrect evaluation can affect the location of the CEP.
The problem is seen most clearly at $T=0$, where the Fermi edge $\phi_F(k)=Z_{\psi,k}(\mu^2-k^2)^{1/2}/g$, the field value at which $E_{\psi,k}=\mu$, is sharp: there the first derivative of the potential is continuous, while the second one jumps (see panels (b) and (d) of Fig.~\ref{fig:lpa_comparison}).\footnote{The quark contribution alone would make also $U_k'$ jump at the edge. In the flow equation~\eqref{Eq:T_flow}, however, the sigma loop prevents this by spreading the jump over a narrow band just outside the edge, where $E_{\sigma,k}\to0$. A minimum lying in this band has a genuinely large $\eta_\sigma$, which is not affected by the treatment described below.}
Derivatives determined from grid points lying on both sides of the Fermi edge give a wrong, even negative, value of the sigma curvature mass squared $M^2_\sigma$ at a minimum lying just inside the Fermi sea.
With a running expansion point, this happens exactly at the value of $k$ at which the expansion point jumps from the global minimum of the broken phase to the one corresponding to the dense phase (see Sec.~\ref{ss:LPAp_run}).
At this stage, the anomalous dimensions, especially $\eta_\sigma$, are computed from an incorrect $M^2_\sigma$, and, in the grid method, the step size of the adaptive solver shrinks until the integration cannot proceed.
For this reason, in both methods we determine the derivatives of the potential only from grid points located on the same side of the Fermi edge as the expansion point: they are computed at the two grid points nearest to the expansion point and linearly interpolated to it, or extrapolated if the expansion point and the edge lie between the same two grid points.
We use finite differences as in the flow of the potential, i.e., second-order formulas with the minimal number of grid points, central where possible and one-sided next to the Fermi edge, so that the masses in the flows of the $Z$ factors are consistent with those in the flow of $U_k$.
In the case of the running minimum, the first derivative is given by the minimum condition, as dictated by the axial Ward identity: $U_k'(\bar\rho_k)=h/\bar\phi_k$ in the grid method, or equivalently $u(\bar s_k)=h$, i.e., $M^2_{\pi,k}=h/\bar s_k$, in the KT scheme, where the bar denotes the extremum.
At $T>0$ the Fermi edge is smeared: the occupation of the quark states changes over $|E_{\psi,k}-\mu|\lesssim5T$.
We treat this band as a Fermi edge only if it is not resolved by the grid, i.e., if it is narrower than a few grid spacings at the current position of the edge, which happens only at temperatures well below $1$~MeV.
If the band is wider, the potential is smooth on the scale of the grid, and the derivatives are computed with central differences.

\subsection{Case of a fixed expansion point}

\begin{figure}[!tb]
  \centering
  \includegraphics[width=0.495\textwidth]{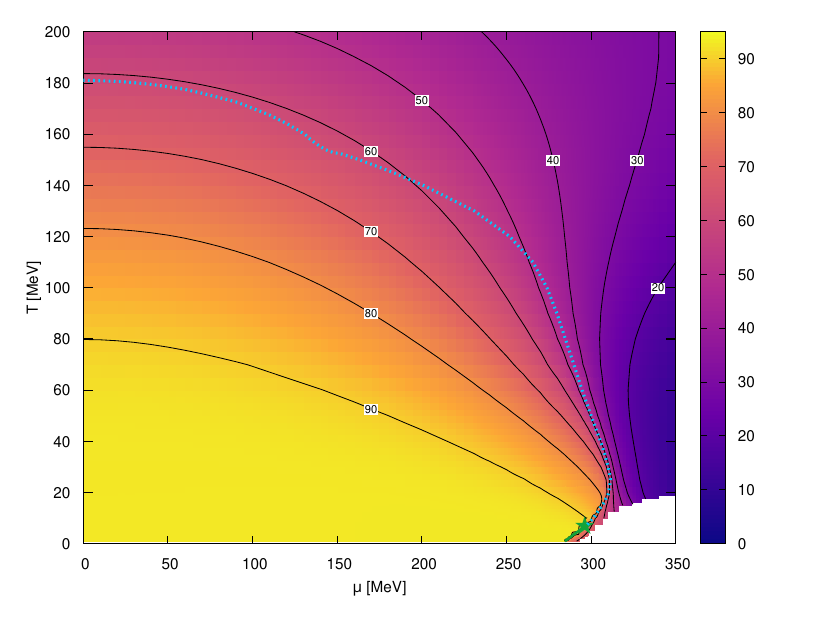}\hfill
  \includegraphics[width=0.495\textwidth]{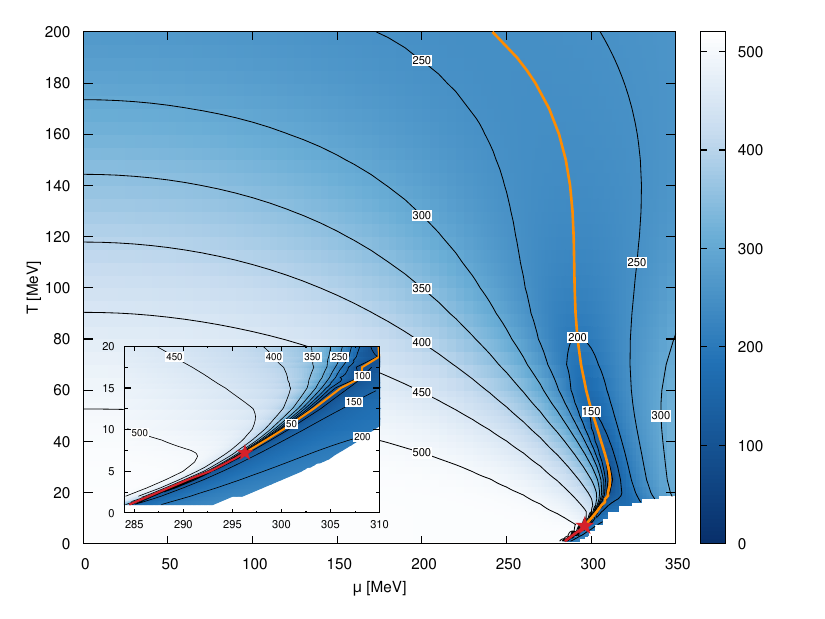}
  \caption{The panels show the same quantities as Fig.~\ref{fig:LPA_k40}, obtained in LPA$'$ at $\Lambda=500$~MeV and $k_{\rm IR}=40$~MeV with the three $Z$ flows evaluated at the fixed expansion point $\phi_\star=96$~MeV (parameters in Table~\ref{tab:params}), only that here the order parameter is the renormalized minimum $\bar\phi_R=Z_\pi^{1/2}\bar\phi$.
    In the white area the flow of $Z_{\sigma,k}$ runs away because the Fermi edge approaches $\phi_\star$: either $Z_{\sigma,k_{\rm IR}}>2$ and the points were excluded (see text), or the flow does not reach $k_{\rm IR}$.
  }
  \label{fig:LPAp_k40}
\end{figure}

By adopting a fixed expansion point $\phi_\star$ (or $\rho_\star = \phi_\star^2/2$), the wave-function renormalizations are evaluated at a constant field value throughout the renormalization group flow.
At finite density, for $\mu > k$, the quark loop induces a sharp structure in the effective potential at the Fermi edge $\phi_F(k)$ defined above.
As the scale $k$ decreases, $\phi_F(k)$ shifts toward larger field values, eventually approaching its infrared (IR) limit, $\mu Z_{\psi,k}/g$.
In the region beyond the first-order phase transition line, this IR limit is close to $\phi_\star$.
Consequently, the flow dynamics at any given point in the $T$--$\mu$ plane are governed by three factors.
\begin{itemize}
\item[(i)] \textit{Proximity of $\phi_F(k)$ to $\phi_\star$:}
  As the Fermi edge approaches $\phi_\star$, the induced structure in the potential reaches $\phi_\star$, causing the anomalous dimension $\eta_\sigma$ to grow via two channels.
  At low temperatures, where the Fermi edge is sharp, the effective potential changes rapidly at $\phi_\star$; the three-point vertex $\mathcal{V}^{(3)}_{\sigma\sigma\sigma}(0,0,0)=3\phi_\star U_k''(\rho_\star)+\phi_\star^3 U_k^{(3)}(\rho_\star)$ increases by almost two orders of magnitude, and $\eta_\sigma$ is predominantly driven by the $\sigma$-bubble diagram.
  When the edge is smeared by the temperature, it is the quark loop that contributes the most to $\eta_\sigma$, through the fractional occupation of quark states at $\phi_\star$, an effect that starts already when the edge is relatively far away.

\item[(ii)] \textit{Scale and duration of the approach:}
  Since $Z_\sigma$ accumulates the anomalous dimension along the flow, it grows the more, the earlier in the flow the Fermi edge approaches $\phi_\star$ and the longer it remains in its vicinity.
  If $\eta_\sigma$ approaches the value $d + 2 = 5$, the regulator factor $1 - \eta_\sigma/(d+2)$ that weights the $\sigma$-loop in the flow equation approaches zero, rendering the flow highly stiff.
  If $\eta_\sigma$ exceeds $5$, this factor changes sign, leading to a breakdown of the integration before reaching $k_{\rm IR}$.

\item[(iii)] \textit{Crossing of $\phi_\star$ by the Fermi edge:}
  If the infrared limit $\mu Z_{\psi,k}/g$ remains below $\phi_\star$, the Fermi edge continuously approaches $\phi_\star$ until the end of the flow, maintaining this enhancement down to $k_{\rm IR}$.
  However, if it lies above $\phi_\star$, the edge would cross $\phi_\star$ during the flow, unless the flow breaks down before.
  Following this crossing, $\phi_\star$ lies inside the Fermi sea, and the enhancement ceases.
  While this crossing is harmless at higher temperatures where the edge is thermally smeared and $Z_\sigma$ remains moderate, at low temperatures, where the edge is sharp and $Z_\sigma$ has already grown significantly, it drives $Z_\sigma$ to very large values, and at the largest values of $\mu$ the flow breaks down before reaching $k_{\rm IR}$.
\end{itemize}
All three factors concern the same quantity, the distance between $\phi_F(k)$ and $\phi_\star$ during the flow.
Together they determine how large $Z_\sigma$ becomes at $k_{\rm IR}$, or whether $k_{\rm IR}$ is reached at all.
This effect is continuous: points immediately adjacent to the excluded region of the phase diagram already exhibit a moderately enhanced $Z_\sigma$, making any specific cutoff (such as $Z_{\sigma,k_{\rm IR}} > 2$) merely a convention.
In Fig.~\ref{fig:LPAp_k40}, which shows the order parameter and the $\sigma$ curvature mass in the $T$--$\mu$ plane obtained with $\phi_\star=96$~MeV and $\Lambda=500$~MeV, the points with $Z_{\sigma,k_{\rm IR}}>2$ and those at which the flow does not reach $k_{\rm IR}$ are left white.
The described problem is specific to the fixed expansion point $\phi_\star$, which the Fermi edge approaches, or sweeps over, irrespective of where the minimum of the potential lies.
Outside the white area of Fig.~\ref{fig:LPAp_k40} the expansion point $\phi_\star$ is not close to a sharp Fermi edge: near the transition line the Fermi edge stays below $\phi_\star$ ($\phi_F\simeq93$~MeV at $k_{\rm IR}$), and where it crosses $\phi_\star$ during the flow, it is smeared by the temperature over many grid spacings.

With the parameter set obtained at $\Lambda=1000$~MeV (last row of Table~\ref{tab:params}), the first-order phase boundary also bends back: the transition occurs at $\mu_{\rm c}=292.6$ and $294.3$~MeV for $T=5$ and $6$~MeV, respectively, and the crossover continues to larger $\mu$ with increasing $T$.
The first-order line ends in a CEP with coordinates given in Table~\ref{tab:CEP}.

\subsection{The case of a running expansion point \label{ss:LPAp_run}}
The problems described above are largely avoided when the wave-function renormalizations are evaluated at the running minimum $\bar\phi_k$ (or $\bar\rho_k=\bar\phi_k^2/2$) of the potential instead of at a fixed $\phi_\star$.
Beyond the first-order line the Fermi edge also approaches the broken-phase minimum, from below, as it approaches $\phi_\star$.
But before it comes close, a second minimum inside the Fermi sea becomes the global one, and the expansion point jumps there.
At the jump the edge is still $7$ to $37$~MeV below the broken-phase minimum, and this distance grows with $\mu$. After the jump the edge keeps moving towards larger fields while the minimum stays, so the two move apart: the expansion point never stays close to the Fermi edge.
The anomalous dimension $\eta_\sigma$ grows only briefly around the jump, and $Z_{\sigma,k_{\rm IR}}$ stays between $1.1$ and $2.8$ beyond the first-order line, decreasing with $\mu$, except within a fraction of an MeV of the transition, where $Z_{\sigma,k_{\rm IR}}$ reaches $4.5$.
The region that was excluded with the fixed $\phi_\star$ thus becomes accessible, and the order parameter can be followed deep into the dense phase.

The first-order line agrees with the one obtained with the fixed $\phi_\star$ to within about $1$~MeV at low temperatures, although the two calculations use different parameter sets and give different sigma masses at $k_{\rm IR}$, $m_\sigma=473.8$ and $500.4$~MeV (Table~\ref{tab:params}).
The critical endpoint lies somewhat higher with the running minimum (see Table~\ref{tab:CEP}).
To separate the effect of the wave-function renormalizations from that of the vacuum sigma mass, we compare each LPA$'$ CEP with the LPA counterpart evaluated with parameters determined with the same vacuum $m_\sigma$.
Relative to the corresponding LPA CEP listed in Table~\ref{tab:CEP}, the LPA$'$ CEP lies lower in temperature by about $5$~MeV with the fixed $\phi_\star$ and $7$~MeV with the running minimum.
This shift is caused by the wave-function renormalizations.

\begin{figure}[!t]
  \centering
  \includegraphics[width=0.495\textwidth]{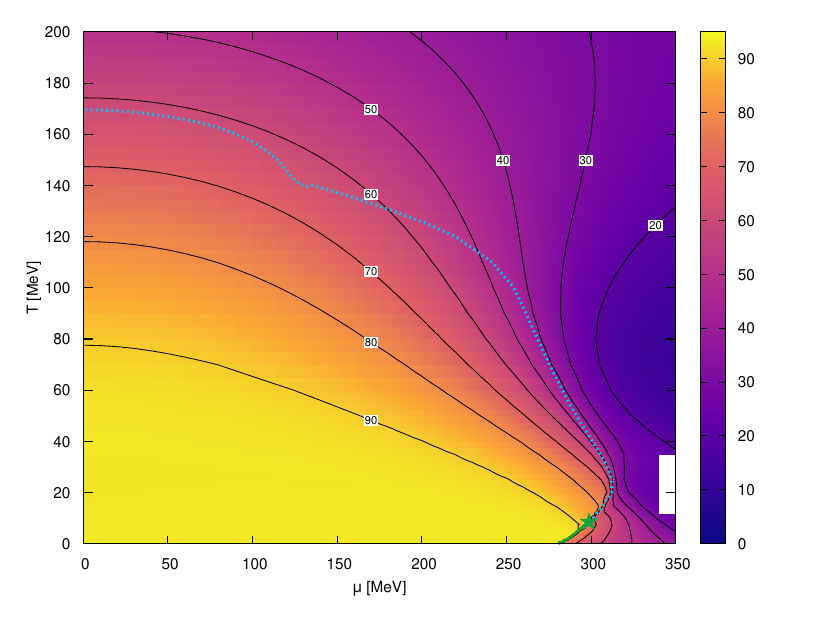}\hfill
  \includegraphics[width=0.495\textwidth]{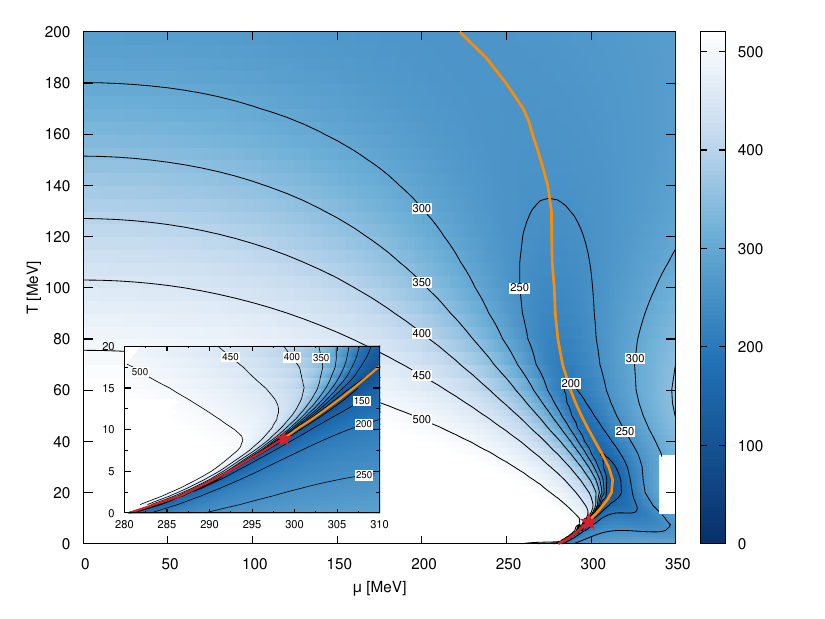}
  \caption{The panels show the quantities of Fig.~\ref{fig:LPAp_k40} obtained in LPA$'$ at $k_{\rm IR}=40$~MeV with the three $Z$ flows evaluated at the running (global) minimum.
    In the white area the flow does not reach $k_{\rm IR}$: the global minimum, and with it the expansion point, jumps early in the flow deep into the Fermi sea, where the flows of the $Z$ factors become too stiff to be integrated.}
  \label{fig:LPAp_run_k40}
\end{figure}

The jump of the expansion point makes the right-hand side of the flow equations of the $Z$'s discontinuous in $k$, and at large $\mu$ this is the source of the remaining difficulty of this approach.
The Fermi edge determines when and where the jump happens: the larger $\mu$, the earlier in the flow the edge approaches the broken-phase minimum, and the deeper inside the Fermi sea the new minimum lies.
At $\mu\simeq350$~MeV and $T\simeq12.5$--$30$~MeV (white area of Fig.~\ref{fig:LPAp_run_k40}) the jump happens already at $k\simeq288$--$306$~MeV and takes the expansion point from $\phi\simeq70$--$86$~MeV to $\phi\simeq15$--$25$~MeV, very deep inside the Fermi sea.
Even when the discontinuity is treated as an event, which is a standard technique for discontinuous ODEs, integrating up to the jump with the old expansion point and restarting with the new one, the flows of the $Z$'s at the new expansion point remain too stiff to be integrated.
At $\mu=340$~MeV, where the new minimum lies at $\phi\simeq19$~MeV, the flows reach $k_{\rm IR}$.

We should mention that the presented LPA$'$ results are exploratory in two respects.
First, $Z_{\sigma,k} = Z_{\pi,k} + \rho\,Y_k$ contains the field explicitly through the coupling $Y_k$, even though $Z_{\pi,k}$ and $Y_k$ are field independent in our truncation.
A moving expansion point therefore contributes a term $Y_k\,d\bar\rho_k/dk$ to the flow of $Z_{\sigma,k}$, which is not included here. 
Second, the $Z$'s are evaluated at the minimum, which depends on the explicit symmetry breaking $h$.
Through the $Z$'s, the potential then depends on $h$ as well, and $1/M_\sigma^2$ is only approximately equal to the chiral susceptibility $\partial\bar\phi/\partial h$.
In contrast, with a fixed $\phi_\star$, the relation is exact.

We also note that our results are obtained with $k_{\rm IR}=40$~MeV.
For $T\lesssim k_{\rm IR}/\pi\simeq13$~MeV the lowest fermionic Matsubara frequency $\pi T$ is smaller than $k_{\rm IR}$, so the thermal fluctuations are not fully integrated out when the flow is stopped, and the results could depend on $k_{\rm IR}$.
We therefore checked the dependence on $k_{\rm IR}$ by repeating the calculation at selected points with $k_{\rm IR}=15$~MeV and parameters refitted to the same vacuum observables, in both LPA and LPA$'$ with $\phi_\star=96$~MeV.
The location of the first-order transition at $T=5$ and $10$~MeV is unchanged within our resolution, and the renormalized minimum $\bar\phi_R$ changes by at most $0.1$~MeV in LPA and $0.2$~MeV in LPA$'$ ($0.7$~MeV in the region where it varies most steeply).
The results obtained with $k_{\rm IR}=40$~MeV are therefore reliable also at these low temperatures.

\section{Conclusions}\label{sec:concl}

We have applied the Kurganov--Tadmor central scheme to the FRG flow of the two-flavor quark-meson model with Litim's regulator, in order to establish a high-resolution benchmark for the standard grid method.
The implementation of the KT scheme for the QM model posed its own challenges.
The standard pointwise estimate of the local propagation speed entering the KT flux had to be made more robust near the nontrivial extrema of the effective potential.
Additionally, the RG-time step had to be controlled through a CFL condition with adaptive step size.
Our implementation turned out to be computationally considerably more expensive than the grid method.

In order to go beyond LPA, we followed Ref.~\cite{Weyrich:2018jid} and computed the flows of $Z_{\pi,k}$ and $Z_{\sigma,k}$ by neglecting $Y_k$ in the three-point vertices, but we retained its contribution through the tadpole diagrams, to see whether such a structural change causes numerical difficulties.
Our flow equations agree with those of Ref.~\cite{Weyrich:2018jid}, with the exception of the tadpole contributions, which were not considered there. 
These flows were solved together with the flow of $Z_{\psi,k}$.

The flow equations of the $Z$ factors were implemented in both codes.
At all points of the $T$--$\mu$ plane where the KT scheme was used as a cross-check, both in LPA and in $\mathrm{LPA}'$, the two methods give virtually identical results, including the discontinuity of the second derivative of the effective potential at the Fermi surface.
Hence, at the resolution used here, the grid method is reliable also in the vicinity of the Fermi surface in both truncations, and it was used to map the corresponding phase diagrams.

The inclusion of the $Z$ factors brought several numerical challenges.
First, the parametrization becomes considerably harder, and in some cases it was not possible to find parameters reproducing the vacuum observables.
With the $Z$ factors evaluated at the running minimum, the target sigma mass cannot be reached at $\Lambda=500$~MeV ($m_\sigma\le473.8$~MeV), and at $\Lambda=1$~GeV no viable parameter set could be found at all, since the constituent quark mass does not exceed about $210$~MeV before the flow breaks down.
At $\Lambda=1$~GeV the model could be parametrized only with a fixed expansion point, and then with a sigma mass well above the target value.
Additionally, we could not reproduce the $k$ dependence of $Z_{\pi,k}$ and $Z_{\sigma,k}$ reported in Ref.~\cite{Weyrich:2018jid} for $T=\mu=0$, which prevented a direct comparison with those results.
Second, at finite density the Fermi edge interferes with the expansion point of the $Z$ flows.
With a fixed expansion point, the approach of the Fermi edge drives $\eta_\sigma$ towards $d+2$, so that $Z_\sigma$ becomes very large, or the flow breaks down, in a region of the $T$--$\mu$ plane beyond the first-order line.
With the running minimum this region becomes accessible, but at large $\mu$ the minimum, and with it the expansion point, jumps early in the flow deep into the Fermi sea, where the flows of the $Z$ factors become too stiff to be integrated.
Third, the derivatives of the potential entering the flows of the $Z$ factors have to be computed from grid points lying on the same side of the Fermi edge as the expansion point.
  Failure to do so leads to incorrect curvature masses at the expansion point, and to a wrong location of the CEP.

In $\mathrm{LPA}'$, we found that the backbending of the first-order transition line persists in all cases studied.
Thus, within the truncation considered here, we do not confirm the absence of backbending reported in Ref.~\cite{Weyrich:2018jid}.
At low temperatures the first-order lines obtained with the two choices of the expansion point agree to within about $1$~MeV, and the critical endpoints differ by only about $2$~MeV in temperature. Compared with LPA at the same sigma mass, the wave-function renormalizations lower $T_{\rm CEP}$ by about $5$--$7$~MeV.

The $\mathrm{LPA}'$ truncation used here could be improved in several respects.
One could use a regulator that does not break the $\mathrm{O}(N)$ symmetry in the mesonic sector, contrary to the one proportional to $Z_{\sigma,k}$ and $Z_{\pi,k}$ used here.
Since the term $Y_k$ already introduces a field dependence of the wave function renormalization, it would be consistent to include also the field derivative $Z'_{\phi,k}$, as well as the contribution $Y_k\,\partial_k\rho_\star$ arising from a running expansion point.
At finite temperature and density the Lorentz symmetry is broken, which could be accounted for by separate wave function renormalizations for the temporal and spatial parts of the kinetic terms.
Finally, the expansion-point problems encountered here could be avoided altogether by solving for the field-dependent functions $Z_{\phi,k}(\rho)$ and $Y_k(\rho)$ together with $U_k(\rho)$, as was done for the $\mathrm{O}(N)$ model in low spatial dimensions in Ref.~\cite{Chlebicki2023}.
For this latter task, a high-resolution scheme, such as the KT method, may again prove useful.

\section*{Acknowledgements}
This work was supported from the Hungarian National Research, Development and Innovation Fund (NRDI Fund) under Project No. K138277.
The authors used the Claude Opus 5.5 and Fable 5.0 models to assist with code development, data generation, and language editing. The resulting code, data and text were reviewed and verified by the authors.

\appendix
\section{Flow of the two-point functions}
\label{App:2point}
In this appendix we give Eq.~\eqref{Eq:Gamma2_flow} explicitly in momentum space, together with the resulting flows of the mesonic two-point functions $\Gamma^{(2)}_{k,\sigma\sigma}$ and $\Gamma^{(2)}_{k,\pi\pi}$ and the vertices entering them, from which $Z_{\sigma,k}$ and $Z_{\pi,k}$ are obtained through Eq.~\eqref{Eq:Zb_def}.
We recall for the sake of completeness that the flow of the quark two-point function, from which $Z_{\psi,k}$ is obtained via Eq.~\eqref{Eq:Zf_def}, is given by the bubble diagram with one quark and one meson propagator:
\begin{equation}
\partial_k \Gamma^{(2)}_{k,\bar\psi\psi}(P) = -\tilde\partial_k \int_Q G_{k}^{ab}(Q-P)\, \Gamma^{(3)}_{k,\bar\psi\psi \phi_a}\, G^R_{\psi,k}(Q)\, \Gamma^{(3)}_{k,\bar\psi\psi \phi_b}.
\label{Eq:Gqq_flow}
\end{equation}
The mesonic field index $a=1,\dots,N$ is such that $\phi_1=\sigma$ and $\phi_{a}=\pi_{a-1}$ for $a\ge2$.
The quark--meson vertices following from the Yukawa term, $\Gamma^{(3)}_{k,\bar\psi\psi \sigma}:=\Gamma^{(3)}_{k,\bar\psi\psi \phi_1}=g$ and $\Gamma^{(3)}_{k,\bar\psi\psi \pi}:=\Gamma^{(3)}_{k,\bar\psi\psi \phi_a}=-\ri g\,\gamma_5\tau_{a-1}$ for $a\ge2$, are momentum independent.
At finite temperature the loop integral is $\int_Q=T\sum_n\int\! d^dq/(2\pi)^d$.

\bigskip
On a homogeneous background, translation invariance implies momentum conservation at each vertex and propagator.
For two mesonic external legs, Eq.~\eqref{Eq:Gamma2_flow} then reads
\begin{eqnarray}
\partial_k \Gamma^{(2)}_{k,ab}(P,-P) &=& \frac{1}{2}\tilde\partial_k \int_Q G_{k}^{mn}(Q)\, \Gamma^{(4)}_{k,nabm}(Q,P,-P,-Q) \nonumber\\
&&-\frac{1}{2}\tilde\partial_k \int_Q G_{k}^{mm'}(Q)\, \Gamma^{(3)}_{k,m'an'}(Q,P,-P-Q)\, G_{k}^{n'n}(Q+P)\, \Gamma^{(3)}_{k,nbm}(Q+P,-P,-Q)\nonumber\\
&&+\tilde\partial_k \int_Q \textrm{tr}\Big[G^R_{\psi,k}(Q)\,\Gamma^{(3)}_{k,\bar\psi\psi \phi_a}\,G^R_{\psi,k}(Q+P)\,\Gamma^{(3)}_{k,\bar\psi\psi \phi_b}\Big],
\label{Eq:2point_flow_mom}
\end{eqnarray}
where summation over repeated indices is implied.
There is no quark tadpole, since the Yukawa term is linear in the meson fields.
The first two terms in Eq.~\eqref{Eq:2point_flow_mom}, which is the explicit form of Eq.~\eqref{Eq:Gamma2_flow}, are the mesonic tadpole and bubble contributions, the last one is the quark loop.
Here $G^{mn}_{k}(Q)=\big[\Gamma_{k}^{(2)}(Q,-Q) +  {\bf R}^{\rm B}_{k}(Q)\big]^{-1}_{mn}$ and $G^R_{\psi,k}=\big(\Gamma^{(2)}_{k,\bar\psi\psi}+R^{\rm F}_k\big)^{-1}$ are the regularized meson and quark propagators, and the trace is over Dirac, color and flavor indices.
Owing to the Grassmann nature of the quark fields, the quark loop enters with the opposite sign and without the factor $1/2$ of the bosonic bubble.

On the homogeneous background $\phi_a=\phi\,\delta_{a1}$, $\rho=\phi^2/2$ (where the delta function of momentum conservation was omitted), the ansatz \eqref{Eq:ansatz} gives
\begin{eqnarray}
\Gamma_{k,ab}^{(2)}(P,-P) &=& M^2_{k,ab}(\rho) + \delta_{ab}\, P^2 Z_{\phi,k} + \delta_{a1} \delta_{b1}\, P^2 \rho\, Y_k ,
\label{Eq:2point_hom}\\
\Gamma_{k,abc}^{(3)}(P_1,P_2,P_3) &=&
\phi\, U''_k(\rho)\big(\delta_{a1} \delta_{bc} + \delta_{b1} \delta_{ac} + \delta_{c1} \delta_{ab} \big) + U'''_k(\rho)\,\phi^3\, \delta_{a1}\delta_{b1} \delta_{c1}
\nonumber \\
&&+\frac{\phi\, Y_k}{2}\Big( (P_1+P_2)^2\, \delta_{ab} \delta_{c1} + (P_1+P_3)^2\, \delta_{ac} \delta_{b1}
 + (P_2+P_3)^2\, \delta_{bc} \delta_{a1} \Big),
\label{Eq:3point_hom}\\
\Gamma_{k,abcd}^{(4)}(P_1,P_2,P_3,P_4) &=&
U''_k(\rho) \big(\delta_{ab} \delta_{cd} + \delta_{ac} \delta_{bd} + \delta_{ad} \delta_{bc} \big)
  + U^{(4)}_k(\rho)\, \phi^4\, \delta_{a1}\delta_{b1}\delta_{c1}\delta_{d1}\nonumber\\
  && + U'''_k(\rho)\,\phi^2 \big(\delta_{ab}\delta_{c1}\delta_{d1} + \delta_{ac}\delta_{b1}\delta_{d1} + \delta_{ad}\delta_{b1}\delta_{c1} +
 \delta_{bc}\delta_{a1}\delta_{d1} + \delta_{bd}\delta_{a1}\delta_{c1} + \delta_{cd}\delta_{a1}\delta_{b1}\big)\nonumber \\
  &&+\frac{Y_k}{2}\Big((P_1+P_2)^2\, \delta_{ab} \delta_{cd} + (P_1+P_3)^2\, \delta_{ac} \delta_{bd}
   + (P_2+P_3)^2\, \delta_{ad} \delta_{bc}\Big),
\label{Eq:4point_hom}
\end{eqnarray}
where $M^2_{k,ab}(\rho)$ is the curvature mass matrix of Eq.~\eqref{Eq:hat_M2}.
Equation~\eqref{Eq:2point_hom} shows that the coefficient of $P^2$ in the $a=b\ge2$ component of $\Gamma^{(2)}_k$ is $Z_{\pi,k}=Z_{\phi,k}$, while in the $a=b=1$ component it is $Z_{\sigma,k}=Z_{\phi,k}+\rho Y_k$, in accordance with Eq.~\eqref{Eq:Zb_def} when evaluated at $\rho=\rho_\star$.
With a regulator diagonal in field space, the meson propagator matrix is diagonal as well:
\begin{equation}
G^{mn}_{k}(Q) = G^R_{\sigma,k}(Q)\,\delta_{m1}\delta_{n1} + G^R_{\pi,k}(Q)\,\big(\delta_{mn}-\delta_{m1}\delta_{n1}\big),
\end{equation}
where the renormalized propagators are given by
\bea
G^R_{\sigma/\pi,k}(Q)= \frac{Z^{-1}_{\sigma/\pi,k}}{\omega_n^2+\vec{q}^{\,2}_R + M^2_{\sigma/\pi,k}/Z_{\sigma/\pi,k}},
\eea
with $Q=(\ri \omega_n,\vec{q})$ and the regularized spatial momentum squared $\vec{q}^{\,2}_R=\vec{q}^{\,2} + \big(k^2-\vec{q}^2\big)\Theta\big(k^2-\vec{q}^2\big)$.

\paragraph{Sigma two-point function}
Setting $a=b=1$ in Eq.~\eqref{Eq:2point_flow_mom} one obtains
\begin{eqnarray}
\partial_k \Gamma^{(2)}_{k,\sigma\sigma}(P,-P) &=&
\frac{1}{2}\tilde\partial_k \int_Q
\Big[ G^R_{\sigma,k}(Q)\, \mathcal{V}^{(4)}_{\sigma\sigma\sigma\sigma}(Q,P,-P,-Q)
  + N_\pi\, G^R_{\pi,k}(Q)\, \mathcal{V}^{(4)}_{\pi\sigma\sigma\pi}(Q,P,-P,-Q) \Big]\nonumber\\
&&-\frac{1}{2}\tilde\partial_k \int_Q
  G^R_{\sigma,k}(Q)\, \mathcal{V}^{(3)}_{\sigma\sigma\sigma}(Q,P,-P-Q)\,
  G^R_{\sigma,k}(Q+P)\, \mathcal{V}^{(3)}_{\sigma\sigma\sigma}(Q+P,-P,-Q)\nonumber\\
&&-\frac{N_\pi}{2}\tilde\partial_k \int_Q
  G^R_{\pi,k}(Q)\, \mathcal{V}^{(3)}_{\pi\sigma\pi}(Q,P,-P-Q)\,
  G^R_{\pi,k}(Q+P)\, \mathcal{V}^{(3)}_{\pi\sigma\pi}(Q+P,-P,-Q)\nonumber\\
&&+\tilde\partial_k \int_Q \textrm{tr}\Big[G^R_{\psi,k}(Q)\,\Gamma^{(3)}_{k,\bar\psi\psi\sigma}\,G^R_{\psi,k}(Q+P)\,\Gamma^{(3)}_{k,\bar\psi\psi\sigma}\Big],
\label{Eq:Gss_flow}
\end{eqnarray}
with $N_\pi=N-1$ and $\mathcal{V}^{(n)}$ the bosonic vertices carrying specific field labels
\begin{subequations}
\begin{eqnarray}
  \mathcal{V}^{(4)}_{\sigma\sigma\sigma\sigma}(Q,P,-P,-Q) &=& 3U''_k(\rho) + 6 U'''_k(\rho)\,\phi^2 + U^{(4)}_k(\rho)\, \phi^4 + (Q+P)^2\, Y_k,\\
  \mathcal{V}^{(4)}_{\pi\sigma\sigma\pi}(P_1,P_2,P_3,P_4) &=& U''_k(\rho) + U'''_k(\rho)\,\phi^2 + \frac{Y_k}{2}(P_2+P_3)^2,
  \label{Eq:V4_pssp}\\
  \mathcal{V}^{(3)}_{\sigma\sigma\sigma}(Q,P,-P-Q) &=&\mathcal{V}^{(3)}_{\sigma\sigma\sigma}(Q+P,-P,-Q) \nonumber \\
  &=& 3 \phi\, U''_k(\rho) + U'''_k(\rho)\,\phi^3 +\frac{\phi\, Y_k}{2}\Big[(Q+P)^2 + P^2 + Q^2 \Big],\\
  \mathcal{V}^{(3)}_{\pi\sigma\pi}(P_1,P_2,P_3) &=&  \phi\, U''_k(\rho) + \frac{\phi}{2}\, P_2^2\,  Y_k ,
  \label{Eq:V3_psp}
\end{eqnarray}
\end{subequations}
where in $\mathcal{V}^{(4)}_{\sigma\sigma\sigma\sigma}$ the symmetry of the integral under $Q\to -Q$ was used.
The vertices are invariant under simultaneous permutations of their field labels and momentum arguments.
The momentum dependence of $\mathcal{V}^{(3)}_{\pi\sigma\pi}$ is given by the momentum of its $\sigma$ leg, and that of $\mathcal{V}^{(4)}_{\pi\sigma\sigma\pi}$ by the total momentum of its $\sigma\sigma$ (or $\pi\pi$) pair, which vanishes in the tadpole.
Only the term proportional to $Y_k$ in $\mathcal{V}^{(4)}_{\sigma\sigma\sigma\sigma}$ depends on the external momentum $P$, hence only this term contributes to $\partial_k Z_{\sigma,k}$ from the tadpole.

\paragraph{Pion two-point function}
Setting $a=b=2$ in Eq.~\eqref{Eq:2point_flow_mom} one obtains
\begin{eqnarray}
\partial_k \Gamma^{(2)}_{k,\pi\pi}(P,-P) &=&
\frac{1}{2}\tilde\partial_k \int_Q
\Big[ G^R_{\pi,k}(Q)\, \mathcal{V}^{(4)}_{\pi\pi\pi\pi}(Q,P,-P,-Q)
  + G^R_{\sigma,k}(Q)\, \mathcal{V}^{(4)}_{\sigma\pi\pi\sigma}(Q,P,-P,-Q) \Big]\nonumber\\
&&-\tilde\partial_k \int_Q
  G^R_{\sigma,k}(Q)\, \mathcal{V}^{(3)}_{\sigma\pi\pi}(Q,P,-P-Q)\,
  G^R_{\pi,k}(Q+P)\, \mathcal{V}^{(3)}_{\pi\pi\sigma}(Q+P,-P,-Q)\nonumber\\
&&+\tilde\partial_k \int_Q \textrm{tr}\Big[G^R_{\psi,k}(Q)\,\Gamma^{(3)}_{k,\bar\psi\psi\pi}\,G^R_{\psi,k}(Q+P)\,\Gamma^{(3)}_{k,\bar\psi\psi\pi}\Big],
\label{Eq:Gpp_flow}
\end{eqnarray}
where the two orderings of the $\sigma$ and $\pi$ propagators in the bubble give equal contributions after the shift $Q\to -Q-P$, which removed the factor $1/2$.
The new vertex compared to Eq.~\eqref{Eq:Gss_flow} is
\begin{equation}
  \mathcal{V}^{(4)}_{\pi\pi\pi\pi}(Q,P,-P,-Q) = (N+1)\, U''_k(\rho) + (Q+P)^2\, Y_k,
\end{equation}
where the sum over the pion components in the loop is included and the symmetry under $Q\to -Q$ was used.
The vertices $\mathcal{V}^{(4)}_{\sigma\pi\pi\sigma}$, $\mathcal{V}^{(3)}_{\sigma\pi\pi}$, and $\mathcal{V}^{(3)}_{\pi\pi\sigma}$ follow from Eqs.~\eqref{Eq:V4_pssp} and \eqref{Eq:V3_psp} by permutation; in particular, the $\sigma\pi\pi$ vertices in the bubble depend on the momentum $Q$ of the $\sigma$ propagator attached to them.
Here as well, only the term proportional to $Y_k$ in $\mathcal{V}^{(4)}_{\pi\pi\pi\pi}$ depends on the external momentum $P$, hence only this term contributes to $\partial_k Z_{\pi,k}$ from the tadpole.

Neglecting $Y_k$ in the three-point vertices of Eqs.~\eqref{Eq:Gss_flow} and \eqref{Eq:Gpp_flow}, expanding the integrals to $\mathcal{O}(\vec p^{\,2})$, performing the Matsubara sums, and then applying the projection \eqref{Eq:Zb_def}, one obtains the flow equations \eqref{Eq:Zsigma_flow} and \eqref{Eq:Zpi_flow}.
We show how to evaluate the contribution of the fermions in Appendix~\ref{app:example}.

\section{Fermionic contribution to $\partial_k Z_{\pi,k}$ and $\partial_k Z_{\sigma,k}$}
\label{app:example}
In this appendix we sketch the method used to obtain the $\mathcal{O}(\vec{p}^{\,2})$ contribution of the integrals appearing in the flow of the two-point functions, which determines the flow of the wave function renormalization factors.
For illustration we use the fermionic contribution to the flow of the bosonic two-point function.

From Eqs.~\eqref{Eq:Gss_flow} and \eqref{Eq:Gpp_flow} we see that we need the expansion of the integral
\be
I_a(P) = \tilde\partial_k \int_Q \textrm{tr}_{\rm D}\Big[G^R_{\psi,k}(Q)\,A_a\,G^R_{\psi,k}(Q+P)\,A_a\Big],
\label{Eq:Ia_def}
\ee
where $\textrm{tr}_{\rm D}$ denotes the trace over Dirac indices only, and $A_1=\mathbb{1}$, $A_2=\gamma_5$.
Performing the color and flavor traces, the quark loops in Eqs.~\eqref{Eq:Gss_flow} and \eqref{Eq:Gpp_flow} are given by $g^2N_cN_f\,I_1(P)$ and $-g^2N_cN_f\,I_2(P)$, respectively, where the minus sign comes from the factor $(-\ri)^2$ of the pion--quark vertices.
In our convention, the quark propagator reads
\be
G^R_{\psi,k}(Q)=\frac{Z^{-1}_{\psi,k}}{\tilde Q_0\gamma^0 + \vec{\slashed{q}}_R + m_{q,k}},
\ee
where $\tilde Q_0=\ri \nu_n -\mu$, $m_{q,k}=M_q/Z_{\psi,k}$ and $\vec{\slashed{q}}_R=\vec{\slashed{q}}\big[1+r_k^F\big(\vec{q}^{\,2}/k^2\big)\big]$.

The operation $\tilde\partial_k$ has to be taken before performing the integral.
Since the external momentum is bosonic, one can set $P^0=0$, as required by the projection \eqref{Eq:Zb_def}, so that $L^0=Q^0$, where $L=Q+P$.
Then, exploiting the symmetry of the integral under $\vec{q}\to-\vec{q}-\vec{p}$, one can write inside the integral $\tilde\partial_k\, \textrm{tr}_{\rm D}\Big[G^R_{\psi,k}(Q)\,A_a\,G^R_{\psi,k}(L)\,A_a\Big] \to -2\,\textrm{tr}_{\rm D}\Big[G^R_{\psi,k}(Q) \partial_k R_k^F(Q) G^R_{\psi,k}(Q) A_a \,G^R_{\psi,k}(L)\,A_a\Big]$.
The derivative of $R_k^F(Q)=Z_{\psi,k}\vec{\slashed{q}} r_k^F(\vec{q}^{\,2}/k^2)$ is $\partial_k R^F_k(Q)=Z_{\psi,k}\frac{\vec{\slashed{q}}}{q}\Big[1-\eta_\psi\Big(1-\frac{q}{k}\Big)\Big] \Theta(k^2-q^2)$, where $q=|\vec{q}|$.

One can then perform the Dirac trace, exploiting $\vec{q}_R^{\,2}=k^2$ in the denominator.
To $\mathcal{O}(\vec{p}^{\,2})$ the integrand of Eq.~\eqref{Eq:Ia_def} becomes $-2 Z^{-2}_{\psi,k}\,q^{-1}\big[1-\eta_\psi\big(1-q/k\big)\big]\Theta(k^2-q^2)\,T_{\sigma/\pi}$, with
\bea
T_\pi&=&-4\big(\vec{q}\cdot(\vec{l}_R-\vec{q}_R)\big)\big(\Delta_k^2(Q)-2\vec{q}_R^{\,2}\Delta_k^3(Q)\big) + 4 \vec{q}\cdot\vec{q}_R\Delta_k(Q)\Delta(L) + \mathcal{O}(|\vec{p}|^3),\\
T_\sigma&=& -T_\pi + 16 m_{q,k}^2 \vec{q}\cdot\vec{q}_R \Delta_k^2(Q)\Delta(L),
\eea
where $\Delta_k(Q)=(m_{q,k}^2+k^2-\tilde Q_0^2)^{-1}$ and $\Delta(L)=(m_{q,k}^2+\vec{l}^{\,2}_R-\tilde Q_0^2)^{-1}$.
Here we exploited that $\vec{q}\cdot(\vec{l}_R-\vec{q}_R)$ is $\mathcal{O}(\vec{p}^{\,2})$, which can be seen by expanding $\vec{q}\cdot\vec{l}_R$ for small $|\vec{p}|$:
\be
\vec{q}\cdot\vec{l}_R = \vec{q}\cdot\vec{q}_R
+ (\vec{q}\cdot \vec{p})^2 \Big[
  \delta(k^2-q^2) + \frac{k}{2 q^3} \Theta(k^2-q^2)
  \Big]
-\vec{p}^{\,2}\frac{k}{2 q} \Theta(k^2-q^2) 
+\mathcal{O}(|\vec{p}|^3).
\ee
In writing this expression we dropped terms of the form $(k-q)\delta(k^2-q^2)$ which cannot contribute under the integral and terms containing $\Theta(q^2-k^2)$ which cannot contribute due to $\Theta(k^2-q^2)$ coming from $\partial_k R^F_k(Q)$.
For this reason there is no linear term in $\vec{p}$.
We also exploited the relation $(k-q) \delta'(k^2-q^2)=-\frac{1}{2k}\delta(k^2-q^2)$ from distribution theory.

In the last term in $T_\pi$ and $T_\sigma$ one has to expand $\Delta(L)$ for small $|\vec{p}|$.
Since $\vec{l}^{\,2}_R=\vec{l}^{\,2}\big(1+r_k^B(\vec{l}^{\,2}/k^2)\big)$, one can use in these terms the expansion given in Eq.~(A12) of Ref.~\cite{Fejos:2025qlv}.
After these expansions, one has to gather the various terms and factors, and apply the Morris Lemma \cite{Morris:1993qb} for the terms containing the Dirac deltas,
\be
\Theta(x)\delta(x) = \delta(x)\int_0^1 d t\, t = \frac{1}{2} \delta(x),
\ee
and the relation
\be
\int \frac{d^d q}{(2\pi)^d} (\vec{p}\cdot \vec{q})^2 F(q^2) =
\frac{\vec{p}^{\,2}}{d} \int \frac{d^d q}{(2\pi)^d} q^2 F(q^2).
\ee
With this procedure the angular integration, the $q$-integral and the Matsubara sum factorize and they can be easily evaluated individually.


\begin{thebibliography}{99}

\bibitem{Aarts:2023vsf}
  G.~Aarts, \textit{et al.}
  ``Phase Transitions in Particle Physics: Results and Perspectives from Lattice Quantum Chromo-Dynamics,''
  Prog. Part. Nucl. Phys. \textbf{133} (2023), 104070
  doi:10.1016/j.ppnp.2023.104070
  [arXiv:2301.04382 [hep-lat]].

\bibitem{Wetterich:1992yh}
  C.~Wetterich,
  ``Exact evolution equation for the effective potential,''
  Phys. Lett. B \textbf{301}, 90-94 (1993).
  doi:10.1016/0370-2693(93)90726-X
  [arXiv:1710.05815 [hep-th]].

\bibitem{Jungnickel:1995fp}
  D.~U.~Jungnickel and C.~Wetterich,
  ``Effective action for the chiral quark-meson model,''
  Phys. Rev. D \textbf{53}, 5142-5175 (1996)
  doi:10.1103/PhysRevD.53.5142
  [arXiv:hep-ph/9505267 [hep-ph]].

\bibitem{Berges:1997eu}
  J.~Berges, D.~U.~Jungnickel and C.~Wetterich,
  ``Two flavor chiral phase transition from nonperturbative flow equations,''
  Phys. Rev. D \textbf{59}, 034010 (1999)
  doi:10.1103/PhysRevD.59.034010
  [arXiv:hep-ph/9705474 [hep-ph]].

\bibitem{Tripolt:2013jra}
  R.~A.~Tripolt, N.~Strodthoff, L.~von Smekal and J.~Wambach,
  ``Spectral Functions for the Quark-Meson Model Phase Diagram from the Functional Renormalization Group,''
  Phys. Rev. D \textbf{89} (2014) no.3, 034010
  doi:10.1103/PhysRevD.89.034010
  [arXiv:1311.0630 [hep-ph]].

\bibitem{Pawlowski:2014zaa}
  J.~M.~Pawlowski and F.~Rennecke,
  ``Higher order quark-mesonic scattering processes and the phase structure of QCD,''
  Phys. Rev. D \textbf{90}, no.7, 076002 (2014)
  doi:10.1103/PhysRevD.90.076002
  [arXiv:1403.1179 [hep-ph]].

\bibitem{Chen:2021iuo} 
  Y.~r.~Chen, R.~Wen and W.~j.~Fu,
  ``Critical behaviors of the O(4) and Z(2) symmetries in the QCD phase diagram,''
  Phys. Rev. D \textbf{104}, no.5, 054009 (2021)
  doi:10.1103/PhysRevD.104.054009
  [arXiv:2101.08484 [hep-ph]].

\bibitem{Schaefer:1999em}
  B.~J.~Schaefer and H.~J.~Pirner,
  ``Renormalization group flow and equation of state of quarks and mesons,''
  Nucl. Phys. A \textbf{660}, 439-474 (1999)
  doi:10.1016/S0375-9474(99)00409-1
  [arXiv:nucl-th/9903003 [nucl-th]].

\bibitem{Schaefer:2004en}
  B.~J.~Schaefer and J.~Wambach,
  ``The Phase diagram of the quark meson model,''
  Nucl. Phys. A \textbf{757}, 479-492 (2005)
  doi:10.1016/j.nuclphysa.2005.04.012
  [arXiv:nucl-th/0403039 [nucl-th]].

\bibitem{Litim:2001hk}
  D.~F.~Litim and J.~M.~Pawlowski,
  ``Predictive power of renormalization group flows: A Comparison,''
  Phys. Lett. B \textbf{516}, 197-207 (2001)
  doi:10.1016/S0370-2693(01)00922-4
  [arXiv:hep-th/0107020 [hep-th]].

\bibitem{Litim:2001up}
  D.~F.~Litim,
  ``Optimized renormalization group flows,''
  Phys. Rev. D \textbf{64}, 105007 (2001)
  doi:10.1103/PhysRevD.64.105007
  [arXiv:hep-th/0103195 [hep-th]].

\bibitem{Toro2009}
  E.~F.~Toro,
  \textit{Riemann Solvers and Numerical Methods for Fluid Dynamics: A Practical Introduction},
  (3rd ed., Springer, Berlin, Heidelberg, 2009)
  doi:10.1007/978-3-540-49834-6.

\bibitem{Grossi:2019urj} 
  E.~Grossi and N.~Wink,
  ``Resolving phase transitions with discontinuous Galerkin methods,''
  SciPost Phys. Core \textbf{6}, 071 (2023)
  doi:10.21468/SciPostPhysCore.6.4.071
  [arXiv:1903.09503 [hep-th]].

\bibitem{KoenigsteinSeries}
  A.~Koenigstein, M.~J.~Steil, N.~Wink, E.~Grossi, J.~Braun, \textit{et al.},
  ``Numerical fluid dynamics for FRG flow equations:
  Zero-dimensional QFTs as numerical test cases. I. The $O(N)$ model,''
  Phys.\ Rev.\ D \textbf{106}, 065012 (2022)
  doi:10.1103/PhysRevD.106.065012
  [arXiv:2108.02504 [cond-mat.stat-mech]];
  A.~Koenigstein, M.~J.~Steil, N.~Wink, E.~Grossi, and J.~Braun,
  ``... II. Entropy production and irreversibility of RG flows,''
  ibid., 065013
  doi:10.1103/PhysRevD.106.065013
  [arXiv:2108.10085 [cond-mat.stat-mech]];
  M.~J.~Steil and A.~Koenigstein,
  ``... III. Shock and rarefaction waves in RG flows reveal limitations
  of the $N\to\infty$ limit in $O(N)$-type models,''
  ibid., 065014
  doi:10.1103/PhysRevD.106.065014
  [arXiv:2108.04037 [cond-mat.stat-mech]].

\bibitem{Stoll:2026ezu}
  J.~Stoll, N.~Zorbach, L.~Kiefer, F.~Murgana, J.~Braun and D.~H.~Rischke,
  ``Analysis of regulator and cutoff artifacts in the phase diagram of the quark-meson model,''
  [arXiv:2601.23005 [hep-ph]].

\bibitem{Scavenius:2000qd}
  O.~Scavenius, A.~Mocsy, I.~N.~Mishustin and D.~H.~Rischke,
  ``Chiral phase transition within effective models with constituent quarks,''
  Phys. Rev. C \textbf{64} (2001), 045202
  doi:10.1103/PhysRevC.64.045202
  [arXiv:nucl-th/0007030 [nucl-th]].

\bibitem{Tripolt:2017zgc}
  R.~A.~Tripolt, B.~J.~Schaefer, L.~von Smekal and J.~Wambach,
  ``Low-temperature behavior of the quark-meson model,''
  Phys. Rev. D \textbf{97}, no.3, 034022 (2018)
  doi:10.1103/PhysRevD.97.034022
  [arXiv:1709.05991 [hep-ph]].

\bibitem{Zhang:2017icm}
  H.~Zhang, D.~Hou, T.~Kojo and B.~Qin,
  ``Functional renormalization group study of the quark-meson model with $\omega$ meson,''
  Phys. Rev. D \textbf{96}, no.11, 114029 (2017)
  doi:10.1103/PhysRevD.96.114029
  [arXiv:1709.05654 [hep-ph]].

\bibitem{Weyrich:2018jid}
  J.~Weyrich,
  ``Quark and nuclear matter with fluctuations,''
  Ph.D. thesis, Universit{\"a}t Giessen (2018)
  doi:10.22029/jlupub-9777.

\bibitem{Grossi:2021ksl} 
  E.~Grossi, F.~J.~Ihssen, J.~M.~Pawlowski and N.~Wink,
  ``Shocks and quark-meson scatterings at large density,''
  Phys. Rev. D \textbf{104}, no.1, 016028 (2021)
  doi:10.1103/PhysRevD.104.016028
  [arXiv:2102.01602 [hep-ph]].

\bibitem{Otto:2022jzl}
  K.~Otto, C.~Busch and B.~J.~Schaefer,
  ``Regulator scheme dependence of the chiral phase transition at high densities,''
  Phys. Rev. D \textbf{106}, no.9, 094018 (2022)
  doi:10.1103/PhysRevD.106.094018
  [arXiv:2206.13067 [hep-ph]].

\bibitem{Berges:2000ew}
  J.~Berges, N.~Tetradis and C.~Wetterich,
  ``Nonperturbative renormalization flow in quantum field theory and statistical physics,''
  Phys. Rept. \textbf{363}, 223-386 (2002).
  doi:10.1016/S0370-1573(01)00098-9.
  [arXiv:hep-ph/0005122 [hep-ph]].

\bibitem{Wetterich:1991be}
  C.~Wetterich,
  ``The Average action for scalar fields near phase transitions,''
  Z. Phys. C \textbf{57}, 451-470 (1993)
  doi:10.1007/BF01474340.

\bibitem{Litim:2000ci}
  D.~F.~Litim,
  ``Optimization of the exact renormalization group,''
  Phys. Lett. B \textbf{486}, 92-99 (2000)
  doi:10.1016/S0370-2693(00)00748-6
  [arXiv:hep-th/0005245 [hep-th]].

\bibitem{Almasi:2017lso}
  G.~A.~Almasi,
  ``Properties of hot and dense strongly interacting matter,''
  Ph.D. thesis, Technische Universit{\"a}t Darmstadt (2017),
  doi:10.26083/tuprints-00006748.

\bibitem{Tetradis:1992qt}
  N.~Tetradis and C.~Wetterich,
  ``Scale dependence of the average potential around the maximum in $\phi^4$ theories,''
  Nucl. Phys. B \textbf{383}, 197-217 (1992)
  doi:10.1016/0550-3213(92)90676-3.

\bibitem{Friederich:2010zn}
  S.~Friederich, H.~C.~Krahl and C.~Wetterich,
  ``Four-point vertex in the Hubbard model and partial bosonization,''
  Phys. Rev. B \textbf{81} (2010), 235108
  doi:10.1103/PhysRevB.81.235108
  [arXiv:1003.0819 [cond-mat.str-el]].

\bibitem{Tanizaki:2013doa}
  Y.~Tanizaki, G.~Fej{\H{o}}s and T.~Hatsuda,
  ``Fermionic Functional Renormalization Group Approach to Superfluid Phase Transition,''
  PTEP \textbf{2014} (2014) no.4, 043I01
  doi:10.1093/ptep/ptu035
  [arXiv:1310.5800 [cond-mat.quant-gas]].

\bibitem{Godunov1959}
  S.~K. Godunov,
  ``A finite difference method for the computation of discontinuous solutions
  of the equations of fluid dynamics,''
  \textit{Mat. Sb.}, vol.~47, pp.~357--393, 1959.

\bibitem{Kurganov:2000ovy}
  A.~Kurganov and E.~Tadmor,
  ``New High-Resolution Central Schemes for Nonlinear Conservation Laws and Convection{\textendash}Diffusion Equations,''
  J. Comput. Phys. \textbf{160}, 241-282 (2000)
  doi:10.1006/jcph.2000.6459

\bibitem{HLL}
  A. ~Harten, P. D. ~Lax, B. ~van Leer,
  ``On Upstream Differencing and Godunov-Type Schemes for Hyperbolic Conservation Laws'',
  SIAM Rev. 25, 35 (1983)
  doi:10.1137/1025002;

\bibitem{HLLC}
  E. F.~Toro, M. ~Spruce, W. ~Speares,
  ``Restoration of the contact surface in the HLL-Riemann solver'',
  Shock Waves 4, 25 (1994)
  doi:10.1007/BF01414629;

\bibitem{Roe}
  P. L. ~Roe,
  ``Approximate Riemann solvers, parameter vectors, and difference schemes'',
  J. Comput. Phys. 43, 357 (1981)
  doi:10.1016/0021-9991(81)90128-5;

\bibitem{Chertock:2004}
  A.~Chertock, A.~ Kurganov and P.~Rosenau,
  ``On degenerate saturated-diffusion equations with convection,''
  Nonlinearity \textbf{18}, 609 (2005)
  doi:10.1088/0951-7715/18/2/009.

\bibitem{CFL}
  R. ~Courant, K. ~Friedrichs and H. ~Lewy,
  ``{\"U}ber die partiellen Differenzengleichungen der mathematischen Physik'',
  Mathematische Annalen, 100, 32--74 (1928)
  doi:10.1007/BF01448839;

\bibitem{Fejos:2025qlv}
  G.~Fej{\H{o}}s, Z.~Sz{\'e}p and N.~Yamamoto,
  ``Scaling behaviors in active model B+ via the functional renormalization group,''
  Phys. Rev. E \textbf{113} (2026) no.1, 014130
  doi:10.1103/k2qg-hrn1
  [arXiv:2508.12065 [cond-mat.stat-mech]].

\bibitem{Bellac:2011kqa}
  M.~L.~Bellac,
  ``Thermal Field Theory,''
  Cambridge University Press, 2011,
  doi:10.1017/CBO9780511721700.

\bibitem{Cohen:2003kd}
  T.~D.~Cohen,
  ``Functional integrals for QCD at nonzero chemical potential and zero density,''
  Phys. Rev. Lett. \textbf{91}, 222001 (2003)
  doi:10.1103/PhysRevLett.91.222001
  [arXiv:hep-ph/0307089 [hep-ph]].
  
\bibitem{Marko:2014hea}
  G.~Mark{\'o}, U.~Reinosa and Z.~Sz{\'e}p,
  ``Bose-Einstein condensation and Silver Blaze property from the two-loop $\Phi$-derivable approximation,''
  Phys. Rev. D \textbf{90}, no.12, 125021 (2014)
  doi:10.1103/PhysRevD.90.125021
  [arXiv:1410.6998 [hep-ph]].
  
\bibitem{Fu:2015naa}
  W.~j.~Fu and J.~M.~Pawlowski,
  ``Relevance of matter and glue dynamics for baryon number fluctuations,''
  Phys. Rev. D \textbf{92}, no.11, 116006 (2015)
  doi:10.1103/PhysRevD.92.116006
  [arXiv:1508.06504 [hep-ph]].
  
\bibitem{Fu:2019hdw}
  W.~j.~Fu, J.~M.~Pawlowski and F.~Rennecke,
  ``QCD phase structure at finite temperature and density,''
  Phys. Rev. D \textbf{101}, no.5, 054032 (2020)
  doi:10.1103/PhysRevD.101.054032
  [arXiv:1909.02991 [hep-ph]].
  
\bibitem{Chlebicki2023}
  A. Chlebicki and P. Jakubczyk,
  ``Low-temperature behavior of the $O(N)$ models below two dimensions,''
  Phys. Rev. E \textbf{107}, 014121 (2023)
  doi:10.1103/PhysRevE.107.014121
  [arXiv:2206.12924 [cond-mat.stat-mech]].

\bibitem{Morris:1993qb}
  T.~R.~Morris,
  ``The Exact renormalization group and approximate solutions,''
  Int. J. Mod. Phys. A \textbf{9}, 2411-2450 (1994)
  doi:10.1142/S0217751X94000972
  [arXiv:hep-ph/9308265 [hep-ph]].
  
\end{thebibliography}
\end{document}